\documentclass[%
 aip,
 amsmath,amssymb,
 reprint,%
]{revtex4-1}

\usepackage{graphicx}
\usepackage{dcolumn}
\usepackage{bm}

\usepackage[utf8]{inputenc}
\usepackage[T1]{fontenc}
\usepackage{mathptmx}
\usepackage{etoolbox}
\usepackage{amsmath}
\makeatletter
\def\@email#1#2{%
 \endgroup
 \patchcmd{\titleblock@produce}
  {\frontmatter@RRAPformat}
  {\frontmatter@RRAPformat{\produce@RRAP{*#1\href{mailto:#2}{#2}}}\frontmatter@RRAPformat}
  {}{}
}%
\makeatother
\begin{document}

\preprint{AIP/123-QED}

\title[]{Mitigation of noise in Josephson parametric oscillator by injection locking}

\author{Gopika Lakshmi Bhai}
\thanks{Authors to whom correspondence should be addressed: gopika.lakshmibhai@gmail.com and tsai@riken.jp}
\affiliation{%
Graduate School of Science, Tokyo University of Science,1–3 Kagurazaka, Shinjuku, Tokyo 162–0825, Japan
}%

\affiliation{%
Research Institute for Science and Technology, Tokyo University of Science, 1-3 Kagurazaka, Shinjuku-ku, Tokyo 162-8601, Japan
}%

\affiliation{%
RIKEN Center for Quantum Computing (RQC), 2–1 Hirosawa, Wako, Saitama 351–0198, Japan
}%
 
\author{Hiroto Mukai}%
 
\affiliation{%
RIKEN Center for Quantum Computing (RQC), 2–1 Hirosawa, Wako, Saitama 351–0198, Japan
}%

\author{Jaw-Shen Tsai}
\thanks{Authors to whom correspondence should be addressed: gopika.lakshmibhai@gmail.com and tsai@riken.jp}
\affiliation{%
Graduate School of Science, Tokyo University of Science,1–3 Kagurazaka, Shinjuku, Tokyo 162–0825, Japan
}%

\affiliation{%
Research Institute for Science and Technology, Tokyo University of Science, 1-3 Kagurazaka, Shinjuku-ku, Tokyo 162-8601, Japan
}%

\affiliation{%
RIKEN Center for Quantum Computing (RQC), 2–1 Hirosawa, Wako, Saitama 351–0198, Japan
}%

\date{\today}

\begin{abstract}
Injection locking is a well-established technique widely used in optics as well as solid-state devices for efficient suppression of noise. We present the spectroscopic characterization of the effect of the injection-locking signal (ILS) in mitigating the phase noise of a Josephson parametric oscillator (JPO), whose output oscillating phase undergoes indeterministic switching between the bistable states with symmetry $\theta \rightarrow{\theta+\pi}$. With the injection of a weak locking signal, we measure the phase noise power spectral density of the self-sustained oscillator output state for different locking signal strengths. We observed suppression of phase noise by injection locking. As the ILS strength surpasses more than a few photons, the output state stays completely pinned to the locking phase of the ILS, and the random telegraphic noise due to the switching of the states is significantly suppressed.
\end{abstract}

\maketitle


Circuit quantum electron dynamics (c-QED) has emerged as one of the rapidly progressing technology for realizing quantum computers~\cite{schmidt2013circuit,arute2019circuit_suprem,buluta2011natural,krantz2019quantum_oliver}. The highly configurable and on-demand engineered superconducting devices in c-QED provide promising and robust hardware solutions~\cite{ladd2010quantum,frunzio2005fabrication,paik2011observation}. In c-QED, the most ubiquitous source of nonlinearity is a Josephson junction (JJ) which opens up the path to the realm of controlled nonlinear physics in solid-state systems~\cite{vrajitoarea2020quantum_jjnonlinearity,albiez2005direct_jjtunneling,zhou2014high_nonlinearJPA,castellanos2008amplification}. Using Josephson circuitry, a great variety of nonlinear effects have been realized, such as frequency conversion~\cite{abdo2013full_freqconv,zakka2011quantum_freqconv,sirois2015coherent_freqconv}, period-multiplying subharmonic oscillations~\cite{svensson2018period_subharmonic,svensson2017period_harmonic}, multi-photon quantum cat states~\cite{vlastakis2013multiphoton,leghtas2015confining_multi}, Kerr effect~\cite{kirchmair2013observation_kerr,yin2012dynamic_Kerr}, etc. More recently, parametrically driven oscillators with Kerr nonlinearity have gathered tremendous popularity since they can yield a cat state via bifurcation~\cite{goto2016bifurcation_goto,goto2019quantum,wang2019quantum_patricio,grimm2020stabilization_puri_bifur,frattini2022squeezed}. These parametric oscillators operate by modulating the natural resonant circuit frequency $\omega_{\mathrm{s}}$ at approximately twice its frequency $\omega_{\mathrm{p}} \approx 2\omega_{\mathrm{s}}$. When the oscillator is strongly driven parametrically above the threshold at $\omega_{\mathrm{p}}$, it undergoes bifurcation, and new stable states emerge with phase separation of $\pi$~\cite{reid1992effect_yurke,yamamoto2016parametric,lin2014josephson}. Above the threshold, various environmental factors, either quantum or classical origin, cause fluctuations in this nonlinear dynamical systems~\cite{dykman2012fluctuating,wilson2010photon_dynamical,wolinsky1988quantum_noise_JPO,kinsler1991quantum_switching}. A recent study on the noise characteristics of a Josephson parametric oscillator (JPO) shows the increase in the phase noise due to the random switching of the oscillator output phase between the two coexisting bistable states~\cite{bhai2022noise}.

In this letter, we present the experimental investigation of the effect of a weak injection-locking signal (ILS) on resonance with the cavity frequency $\omega_{\mathrm{s}}$ in the mitigation of the phase noise of a JPO operating above the threshold. A number of experiments and theoretical studies have demonstrated injection locking in solid-state systems, e.g., Josephson parametric phase-locked oscillators~\cite{lin2014josephson,krantz2016single,bengtsson2018nondegenerate,svensson2018period_subharmonic}, Josephson photonic devices~\cite{danner2021injection_photonic}, masers~\cite{toth2018maser}, lasers~\cite{buczek1973laser}, etc. Here, we spectroscopically characterize the effect of the ILS on the mitigation of the phase noise in JPO  by measuring the phase noise power spectral density (PSD) for different input strengths of the ILS.

\begin{figure}
\begin{center}
\includegraphics[keepaspectratio]{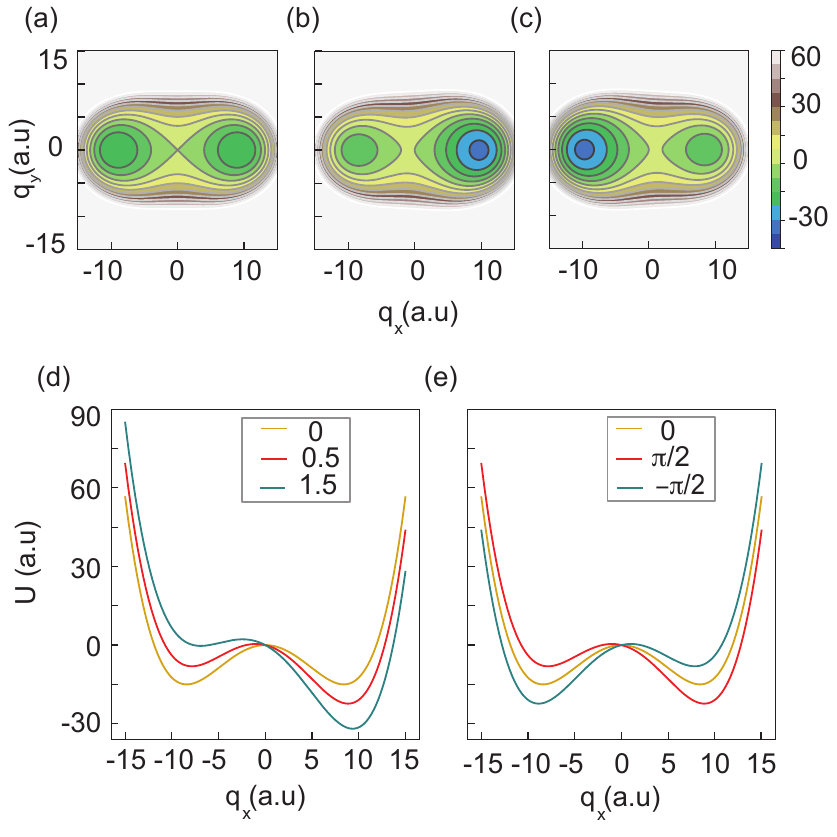} 
\end{center}
\caption{\label{Fig1} 
(a)~Illustration of the effective potential of the JPO in the absence of the ILS. The potential has two symmetrically placed minima. (b)~ A weak ILS is applied, and the phase of ILS is varied, $\theta_{\mathrm{s}} = \pi/2$ (c) $\theta_{\mathrm{s}} = -\pi/2$. (d)~cross section for fixed $\theta_{\mathrm{s}}$ and different ILS strength represented in the mean photon number in the resonator cavity, $N_{\rm{p}}$. (e)~cross section at fixed $N_{\rm{p}}$ and different $\theta_{\mathrm{s}}$.}
\end{figure}

\begin{figure*}
\begin{center}
\includegraphics[keepaspectratio]{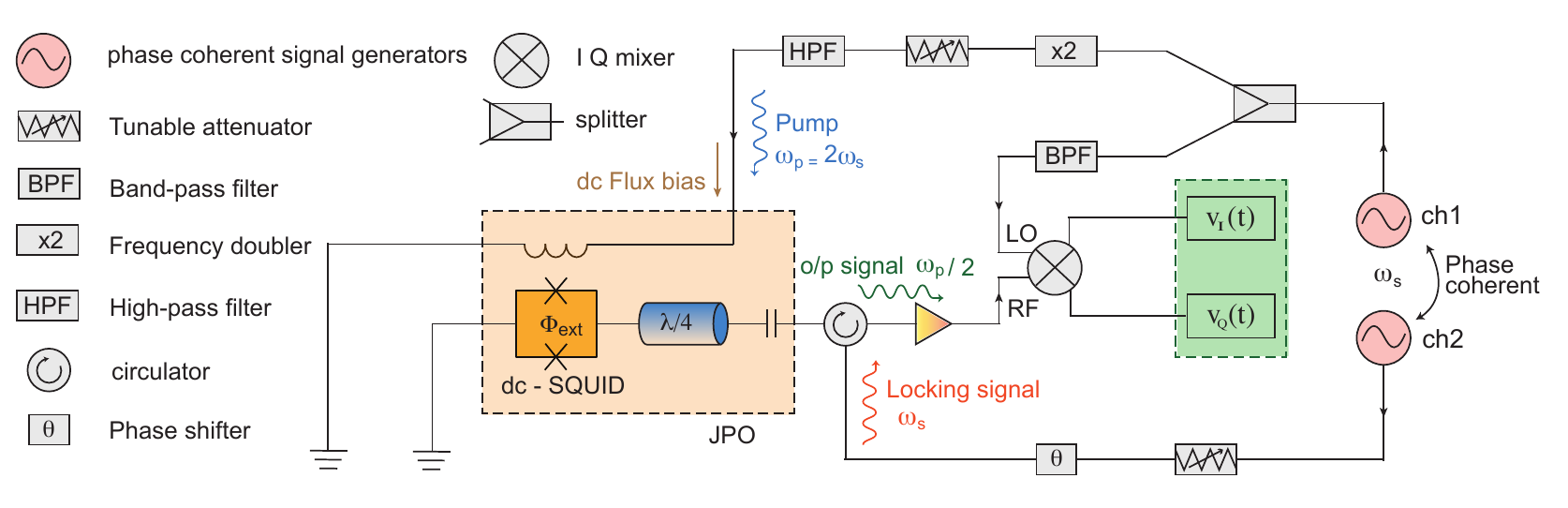} 
\end{center}
\caption{\label{Fig2} 
Schematic of the noise measurement setup used to extract the phase noise spectral density of the JPO.
The dotted brown box indicates the JPO chip based on a quarter wavelength resonator terminated by a dc-SQUID. A tunable attenuator and phase shifter are used to control the power and phase of the signal. The dotted green box represents the data acquisition (DAQ) system that digitizes the unbreakable sequence of I and Q signal voltages for 10 s with a sampling rate of 1MSa/s.}
\end{figure*}
Our JPO device is implemented using a superconducting quarter wavelength coplanar waveguide resonator (CPW) short-circuited to the ground by a dc-SQUID~\cite{yamamoto2008flux} (see Ref.~Supplementary section S1 for device details). By tuning the magnetic flux threading through the dc-SQUID with a dc bias, the resonance frequency of the resonator can be varied~\cite{yamamoto2008flux,roy2016introduction}. An on-chip antenna inductively coupled to the dc-SQUID is used to apply the coherent pump tone $\omega_{\mathrm{p}}$. When the strength of the pump tone $P_{\mathrm{p}}$ exceeds the parametric threshold $P_{\mathrm{th}}$, cavity field amplitude exponentially grows, and the system enters into the oscillating state~\cite{yamamoto2016parametric,krantz2013investigation,krantz2016single,wustmann2019parametric}. The oscillator output states exhibit bistability with two distinct self-oscillating states with well-defined phases with a relative phase shift of $\pi$, called the 0$\pi$ and 1$\pi$ states~\cite{lin2014josephson,reid1992effect_yurke}. Injecting a small signal at the resonance frequency of the oscillator breaks the degeneracy of the system, and the oscillator locks to the phase of the injected signal~\cite{yamamoto2016parametric}. This mechanism of injection locking 
can be understood from the bistable potential~\cite{dykman1998fluctuational_phaseflip} of the JPO described by (see Ref.~\citenum{lin2014josephson} for detailed derivation),
\begin{multline}
U(q_{x}, q_{y}) = \frac{\kappa}{4}\sqrt{\frac{P_{\rm{p}}}{P_{\rm{th}}}}\left(q_{y}^2 - q_{x}^2\right)
\\
- 3\gamma (q_{x}^2 + q_{y}^2)^2 + \sqrt{\kappa_{\rm{ext}}}|E_{\rm{s}}|\left(q_{y}\cos{\theta_{\rm{s}}} - q_{x}\sin{\theta_{\rm{s}}}\right).
\end{multline}

Here $[q_{x}(t) - iq_{y}(t)]e^{-i\omega_{\rm{s}}t} = \langle a\rangle$ is the resonator field, and $\kappa = \kappa_{\rm{ext}} + \kappa_{\rm{int}}$, where
$\kappa_{\rm{ext}}$ and $\kappa_{\rm{int}}$ are external and internal loss rates of the resonator. $\gamma$ denotes the nonlinearity of the Josephson junction~\cite{lin2014josephson}, $|E_{\rm{s}}| = \sqrt{P_{\rm{s}}/\hbar\omega_{\rm{s}}}$ and $\theta_{\rm{s}}$ are the amplitude and phase of ILS. 

Fig.~1 shows the illustration of the potential. In Fig~1(a), we see that the unperturbed bistable potential, in the absence of the ILS, obeys symmetry with two equal minima. When a weak ILS is applied with a phase of either $\pi/2$ or $-\pi/2$, the symmetry is violated, and asymmetry is introduced where the oscillator state locks to the ILS phase as shown in Fig.~1(b) and (c). Due to the lift in the degeneracy, one of the states will be lower in energy which makes it populated with a higher probability. Fig.~1(d) and (e) show the cross-section of the bistable potential $U(q_{x}, 0)$. We note that the potential minima deepen as we increase the ILS strength, as shown in Fig.~1(d). Furthermore, in Fig.~1(e), we see that when we vary the phase of the ILS, the potential tilts according to the injected phase favoring either 0$\pi$ or 1$\pi$ state. 

\begin{figure}
\begin{center}
\includegraphics[keepaspectratio]{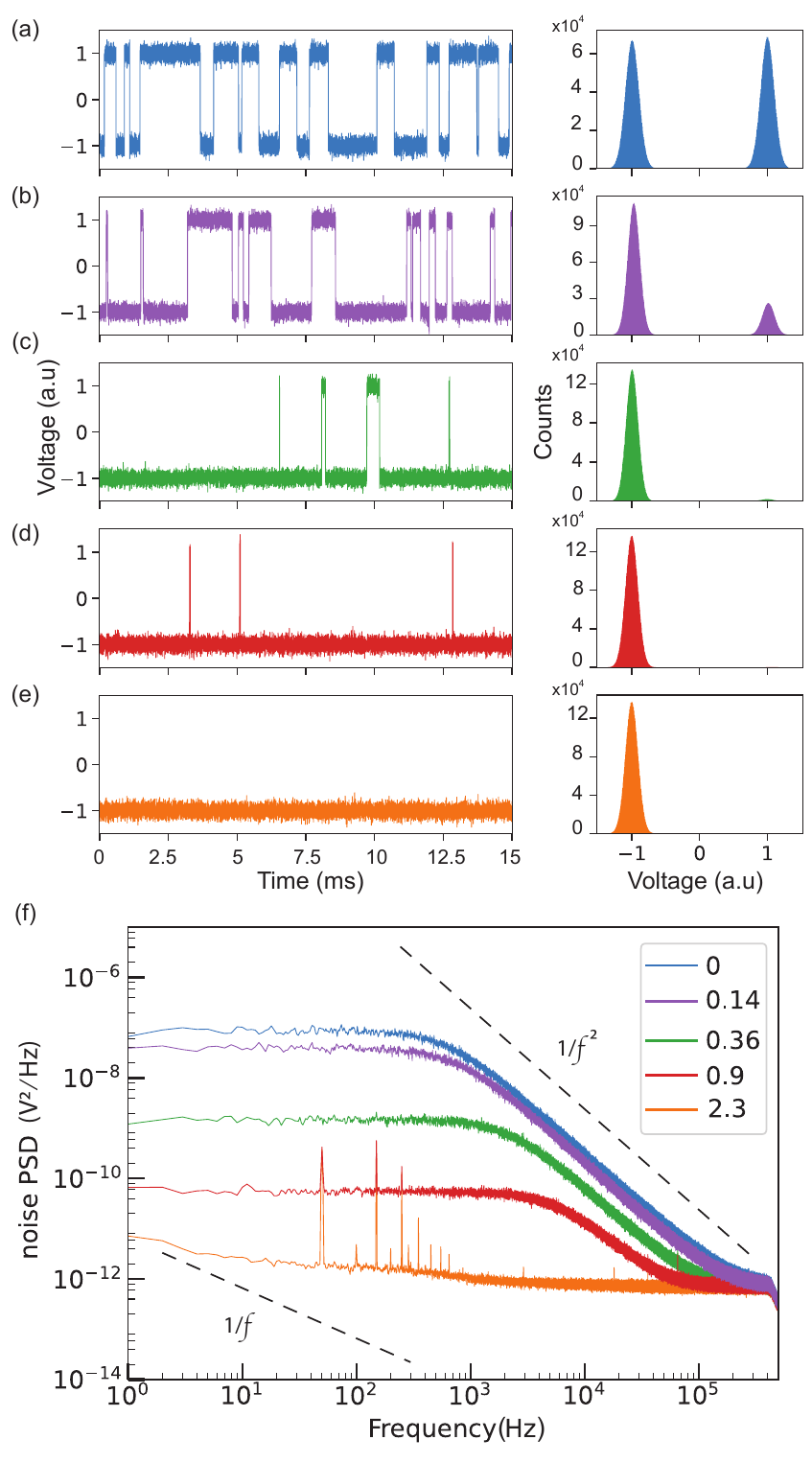} 
\end{center}
\caption{\label{Fig2}
(a-e)~JPO output trajectory in the I quadrature as a function of time cropped from the full time trace for different ILS strengths represented in the cavity photon number $N_{\rm{p}}$. The pump power is fixed at $P_{\rm{p}}$ = -56 dBm and phase of the ILS is fixed at $\theta_{\rm{s}} = -\pi/2$. The right panel shows the histograms of the corresponding time trace on the left using the full time trace.
(f)~Phase noise PSD of the JPO plotted for different ILS strengths. $N_{\rm{p}} = 0$ indicates absence of ILS. The spurs around 50 Hz and above are due to the ac line harmonics~\cite{santoso2012electrical_spur}. $1/f$ and $1/f^{2}$ frequency characteristics are plotted in dotted lines for reference. The corresponding PSDs are color coded with the time traces and histogram plots in (a-e).}
\end{figure}

In order to experimentally analyze the effect of the ILS on the JPO output dynamics and to characterize the phase noise in JPO in the presence of ILS, we adopt the homodyne interferometric measurement setup, as shown in Fig.~2. The JPO device is cooled down to a temperature of 10 mK using a dilution refrigerator (see Ref.~S1 for details of our cryogenic circuit). Two phase-coherent channels of Anapico microwave signal generator feed as the pump tone at $\omega_{\mathrm{p}}$ and injection locking tone at $\omega_{\mathrm{s}}$. The signal from channel-1 is split into two with a microwave splitter to provide one of the signals as the local oscillator. The signal toward the pump port is frequency multiplied to $2\omega_{\mathrm{s}}$ using a frequency doubler. A series of filters are used in the circuit paths to filter out the unwanted frequencies. Through the input port of the JPO, the locking signal from channel-2 is fed through a circulator, and the output from the JPO is amplified using an amplifying chain and is connected to the RF port of the IQ mixer for digitizing the I and Q time traces continuously for 10 seconds with a sampling rate of 1MSa/s. These time traces are further analyzed by constructing a spectral domain noise covariance matrix~\cite{gao2007noise} for extracting noise PSDs in both phase and amplitude quadrature~\cite{bhai2022noise} (Ref.~S2 for the details of the data analysis). 

 
The experimental results are shown in Fig.~3. Fig.~3(a-e) shows the 15~ms time trace cropped from the full 10~s long data set. Fig.~3(a) shows the scaled voltage in I quadrature as a function of time in the absence of the ILS where the signal fluctuates between two states exhibiting random telegraphic noise~\cite{machlup1954noise_RTS,yuzhelevski2000random_RTS}. The right panel shows the histogram of the full 10~s time trace. We see that in the absence of the ILS, the output state shows approximately equal weights corresponding to the 0$\pi$ and 1$\pi$ states. With the injection of the locking signal, the distribution starts to vary, with one state more favorable than the other as the degeneracy of the potential lifts. As the strength of the ILS increases, the output state tends to stay pinned to the locking signal phase, and thus the random telegraphic noise is observed to be significantly reducing. 

Fig.~3(f) shows the noise power spectral density in the phase quadrature for different input locking signal strengths, which is proportional to the cavity photon number. The noise PSDs have a Lorentzian shape which characterizes the random telegraphic noise~\cite{machlup1954noise_RTS,yuzhelevski2000random_RTS}. The noise PSDs are flat at low frequencies and falls off with a $1/f^{2}$ trend and merge with the white noise. As the ILS strength increases, the phase noise effectively suppresses. When the cavity photon number exceeds a few photons, $N_{p} \approx 2.3$, the state stays completely pinned to $0\pi$ state with no switching events observed for 10 s, as shown in Fig.~3(e). At this ILS power, we observe a significant reduction in the phase noise, which shows a 1/f trend at low frequencies and falls off to the white frequency noise. This demonstrates the phase diffusion of the oscillator state due to the fast fluctuation of the phase localized in one of the oscillating states. 

In summary, experimental observation on the phase noise behavior of the injection-locked Josephson parametric oscillator has been presented. The phase noise of the JPO is significantly suppressed with the injection locking. When the ILS strength exceeds a few photons, the output state stays pinned to one of the oscillating states. As a result, we observe a complete suppression of the random telegraphic noise. At this power, the phase diffusion of the oscillating state in a localized state is observed to impart a white noise.
\vspace{0.4cm}

See the supplementary material for more information about the experimental setup and data analysis.
\vspace{0.4cm}

We acknowledge fruitful discussions with T.~Yamamoto, V.~Sudhir, J.~Gao, M.~I.~Dykman, E.~Rubiola, and K.~Koshino. We are grateful to R.~Wang, and P.~Patil for their thoughtful comments. Thanks to K.~Nittoh for support in fabrication. We also thank K.~Kikuchi for the technical support from Keysight. This research work was supported in part by JST CREST (Grant No.~JPMJCR1676 and JPMJCR1775), New Energy and Industrial Technology Development Organization (NEDO)(Grant No.~JPNP16007), and Moonshot R \& D (Grant No.~JPMJMS2067). H. M. was supported by RIKEN SPR project.

\section*{References}
\bibliography{main_ref}

\providecommand{\noopsort}[1]{}\providecommand{\singleletter}[1]{#1}%
\begin{thebibliography}{47}%
\makeatletter
\providecommand \@ifxundefined [1]{%
 \@ifx{#1\undefined}
}%
\providecommand \@ifnum [1]{%
 \ifnum #1\expandafter \@firstoftwo
 \else \expandafter \@secondoftwo
 \fi
}%
\providecommand \@ifx [1]{%
 \ifx #1\expandafter \@firstoftwo
 \else \expandafter \@secondoftwo
 \fi
}%
\providecommand \natexlab [1]{#1}%
\providecommand \enquote  [1]{``#1''}%
\providecommand \bibnamefont  [1]{#1}%
\providecommand \bibfnamefont [1]{#1}%
\providecommand \citenamefont [1]{#1}%
\providecommand \href@noop [0]{\@secondoftwo}%
\providecommand \href [0]{\begingroup \@sanitize@url \@href}%
\providecommand \@href[1]{\@@startlink{#1}\@@href}%
\providecommand \@@href[1]{\endgroup#1\@@endlink}%
\providecommand \@sanitize@url [0]{\catcode `\\12\catcode `\$12\catcode
  `\&12\catcode `\#12\catcode `\^12\catcode `\_12\catcode `\%12\relax}%
\providecommand \@@startlink[1]{}%
\providecommand \@@endlink[0]{}%
\providecommand \url  [0]{\begingroup\@sanitize@url \@url }%
\providecommand \@url [1]{\endgroup\@href {#1}{\urlprefix }}%
\providecommand \urlprefix  [0]{URL }%
\providecommand \Eprint [0]{\href }%
\providecommand \doibase [0]{http://dx.doi.org/}%
\providecommand \selectlanguage [0]{\@gobble}%
\providecommand \bibinfo  [0]{\@secondoftwo}%
\providecommand \bibfield  [0]{\@secondoftwo}%
\providecommand \translation [1]{[#1]}%
\providecommand \BibitemOpen [0]{}%
\providecommand \bibitemStop [0]{}%
\providecommand \bibitemNoStop [0]{.\EOS\space}%
\providecommand \EOS [0]{\spacefactor3000\relax}%
\providecommand \BibitemShut  [1]{\csname bibitem#1\endcsname}%
\let\auto@bib@innerbib\@empty
\bibitem [{\citenamefont {Schmidt}\ and\ \citenamefont
  {Koch}(2013)}]{schmidt2013circuit}%
  \BibitemOpen
  \bibfield  {author} {\bibinfo {author} {\bibfnamefont {S.}~\bibnamefont
  {Schmidt}}\ and\ \bibinfo {author} {\bibfnamefont {J.}~\bibnamefont {Koch}},\
  }\bibfield  {title} {\enquote {\bibinfo {title} {Circuit qed lattices:
  Towards quantum simulation with superconducting circuits},}\ }\href@noop {}
  {\bibfield  {journal} {\bibinfo  {journal} {Annalen der Physik}\ }\textbf
  {\bibinfo {volume} {525}},\ \bibinfo {pages} {395--412} (\bibinfo {year}
  {2013})}\BibitemShut {NoStop}%
\bibitem [{\citenamefont {Arute}\ \emph {et~al.}(2019)\citenamefont {Arute},
  \citenamefont {Arya}, \citenamefont {Babbush}, \citenamefont {Bacon},
  \citenamefont {Bardin}, \citenamefont {Barends}, \citenamefont {Biswas},
  \citenamefont {Boixo}, \citenamefont {Brandao}, \citenamefont {Buell} \emph
  {et~al.}}]{arute2019circuit_suprem}%
  \BibitemOpen
  \bibfield  {author} {\bibinfo {author} {\bibfnamefont {F.}~\bibnamefont
  {Arute}}, \bibinfo {author} {\bibfnamefont {K.}~\bibnamefont {Arya}},
  \bibinfo {author} {\bibfnamefont {R.}~\bibnamefont {Babbush}}, \bibinfo
  {author} {\bibfnamefont {D.}~\bibnamefont {Bacon}}, \bibinfo {author}
  {\bibfnamefont {J.~C.}\ \bibnamefont {Bardin}}, \bibinfo {author}
  {\bibfnamefont {R.}~\bibnamefont {Barends}}, \bibinfo {author} {\bibfnamefont
  {R.}~\bibnamefont {Biswas}}, \bibinfo {author} {\bibfnamefont
  {S.}~\bibnamefont {Boixo}}, \bibinfo {author} {\bibfnamefont {F.~G.}\
  \bibnamefont {Brandao}}, \bibinfo {author} {\bibfnamefont {D.~A.}\
  \bibnamefont {Buell}},  \emph {et~al.},\ }\bibfield  {title} {\enquote
  {\bibinfo {title} {Quantum supremacy using a programmable superconducting
  processor},}\ }\href@noop {} {\bibfield  {journal} {\bibinfo  {journal}
  {Nature}\ }\textbf {\bibinfo {volume} {574}},\ \bibinfo {pages} {505--510}
  (\bibinfo {year} {2019})}\BibitemShut {NoStop}%
\bibitem [{\citenamefont {Buluta}, \citenamefont {Ashhab},\ and\ \citenamefont
  {Nori}(2011)}]{buluta2011natural}%
  \BibitemOpen
  \bibfield  {author} {\bibinfo {author} {\bibfnamefont {I.}~\bibnamefont
  {Buluta}}, \bibinfo {author} {\bibfnamefont {S.}~\bibnamefont {Ashhab}}, \
  and\ \bibinfo {author} {\bibfnamefont {F.}~\bibnamefont {Nori}},\ }\bibfield
  {title} {\enquote {\bibinfo {title} {Natural and artificial atoms for quantum
  computation},}\ }\href@noop {} {\bibfield  {journal} {\bibinfo  {journal}
  {Reports on Progress in Physics}\ }\textbf {\bibinfo {volume} {74}},\
  \bibinfo {pages} {104401} (\bibinfo {year} {2011})}\BibitemShut {NoStop}%
\bibitem [{\citenamefont {Krantz}\ \emph {et~al.}(2019)\citenamefont {Krantz},
  \citenamefont {Kjaergaard}, \citenamefont {Yan}, \citenamefont {Orlando},
  \citenamefont {Gustavsson},\ and\ \citenamefont
  {Oliver}}]{krantz2019quantum_oliver}%
  \BibitemOpen
  \bibfield  {author} {\bibinfo {author} {\bibfnamefont {P.}~\bibnamefont
  {Krantz}}, \bibinfo {author} {\bibfnamefont {M.}~\bibnamefont {Kjaergaard}},
  \bibinfo {author} {\bibfnamefont {F.}~\bibnamefont {Yan}}, \bibinfo {author}
  {\bibfnamefont {T.~P.}\ \bibnamefont {Orlando}}, \bibinfo {author}
  {\bibfnamefont {S.}~\bibnamefont {Gustavsson}}, \ and\ \bibinfo {author}
  {\bibfnamefont {W.~D.}\ \bibnamefont {Oliver}},\ }\bibfield  {title}
  {\enquote {\bibinfo {title} {A quantum engineer's guide to superconducting
  qubits},}\ }\href@noop {} {\bibfield  {journal} {\bibinfo  {journal} {Applied
  Physics Reviews}\ }\textbf {\bibinfo {volume} {6}},\ \bibinfo {pages}
  {021318} (\bibinfo {year} {2019})}\BibitemShut {NoStop}%
\bibitem [{\citenamefont {Ladd}\ \emph {et~al.}(2010)\citenamefont {Ladd},
  \citenamefont {Jelezko}, \citenamefont {Laflamme}, \citenamefont {Nakamura},
  \citenamefont {Monroe},\ and\ \citenamefont {O’Brien}}]{ladd2010quantum}%
  \BibitemOpen
  \bibfield  {author} {\bibinfo {author} {\bibfnamefont {T.~D.}\ \bibnamefont
  {Ladd}}, \bibinfo {author} {\bibfnamefont {F.}~\bibnamefont {Jelezko}},
  \bibinfo {author} {\bibfnamefont {R.}~\bibnamefont {Laflamme}}, \bibinfo
  {author} {\bibfnamefont {Y.}~\bibnamefont {Nakamura}}, \bibinfo {author}
  {\bibfnamefont {C.}~\bibnamefont {Monroe}}, \ and\ \bibinfo {author}
  {\bibfnamefont {J.~L.}\ \bibnamefont {O’Brien}},\ }\bibfield  {title}
  {\enquote {\bibinfo {title} {Quantum computers},}\ }\href@noop {} {\bibfield
  {journal} {\bibinfo  {journal} {nature}\ }\textbf {\bibinfo {volume} {464}},\
  \bibinfo {pages} {45--53} (\bibinfo {year} {2010})}\BibitemShut {NoStop}%
\bibitem [{\citenamefont {Frunzio}\ \emph {et~al.}(2005)\citenamefont
  {Frunzio}, \citenamefont {Wallraff}, \citenamefont {Schuster}, \citenamefont
  {Majer},\ and\ \citenamefont {Schoelkopf}}]{frunzio2005fabrication}%
  \BibitemOpen
  \bibfield  {author} {\bibinfo {author} {\bibfnamefont {L.}~\bibnamefont
  {Frunzio}}, \bibinfo {author} {\bibfnamefont {A.}~\bibnamefont {Wallraff}},
  \bibinfo {author} {\bibfnamefont {D.}~\bibnamefont {Schuster}}, \bibinfo
  {author} {\bibfnamefont {J.}~\bibnamefont {Majer}}, \ and\ \bibinfo {author}
  {\bibfnamefont {R.}~\bibnamefont {Schoelkopf}},\ }\bibfield  {title}
  {\enquote {\bibinfo {title} {Fabrication and characterization of
  superconducting circuit qed devices for quantum computation},}\ }\href@noop
  {} {\bibfield  {journal} {\bibinfo  {journal} {IEEE transactions on applied
  superconductivity}\ }\textbf {\bibinfo {volume} {15}},\ \bibinfo {pages}
  {860--863} (\bibinfo {year} {2005})}\BibitemShut {NoStop}%
\bibitem [{\citenamefont {Paik}\ \emph {et~al.}(2011)\citenamefont {Paik},
  \citenamefont {Schuster}, \citenamefont {Bishop}, \citenamefont {Kirchmair},
  \citenamefont {Catelani}, \citenamefont {Sears}, \citenamefont {Johnson},
  \citenamefont {Reagor}, \citenamefont {Frunzio}, \citenamefont {Glazman}
  \emph {et~al.}}]{paik2011observation}%
  \BibitemOpen
  \bibfield  {author} {\bibinfo {author} {\bibfnamefont {H.}~\bibnamefont
  {Paik}}, \bibinfo {author} {\bibfnamefont {D.~I.}\ \bibnamefont {Schuster}},
  \bibinfo {author} {\bibfnamefont {L.~S.}\ \bibnamefont {Bishop}}, \bibinfo
  {author} {\bibfnamefont {G.}~\bibnamefont {Kirchmair}}, \bibinfo {author}
  {\bibfnamefont {G.}~\bibnamefont {Catelani}}, \bibinfo {author}
  {\bibfnamefont {A.~P.}\ \bibnamefont {Sears}}, \bibinfo {author}
  {\bibfnamefont {B.}~\bibnamefont {Johnson}}, \bibinfo {author} {\bibfnamefont
  {M.}~\bibnamefont {Reagor}}, \bibinfo {author} {\bibfnamefont
  {L.}~\bibnamefont {Frunzio}}, \bibinfo {author} {\bibfnamefont {L.~I.}\
  \bibnamefont {Glazman}},  \emph {et~al.},\ }\bibfield  {title} {\enquote
  {\bibinfo {title} {Observation of high coherence in josephson junction qubits
  measured in a three-dimensional circuit qed architecture},}\ }\href@noop {}
  {\bibfield  {journal} {\bibinfo  {journal} {Physical Review Letters}\
  }\textbf {\bibinfo {volume} {107}},\ \bibinfo {pages} {240501} (\bibinfo
  {year} {2011})}\BibitemShut {NoStop}%
\bibitem [{\citenamefont {Vrajitoarea}\ \emph {et~al.}(2020)\citenamefont
  {Vrajitoarea}, \citenamefont {Huang}, \citenamefont {Groszkowski},
  \citenamefont {Koch},\ and\ \citenamefont
  {Houck}}]{vrajitoarea2020quantum_jjnonlinearity}%
  \BibitemOpen
  \bibfield  {author} {\bibinfo {author} {\bibfnamefont {A.}~\bibnamefont
  {Vrajitoarea}}, \bibinfo {author} {\bibfnamefont {Z.}~\bibnamefont {Huang}},
  \bibinfo {author} {\bibfnamefont {P.}~\bibnamefont {Groszkowski}}, \bibinfo
  {author} {\bibfnamefont {J.}~\bibnamefont {Koch}}, \ and\ \bibinfo {author}
  {\bibfnamefont {A.~A.}\ \bibnamefont {Houck}},\ }\bibfield  {title} {\enquote
  {\bibinfo {title} {Quantum control of an oscillator using a stimulated
  josephson nonlinearity},}\ }\href@noop {} {\bibfield  {journal} {\bibinfo
  {journal} {Nature Physics}\ }\textbf {\bibinfo {volume} {16}},\ \bibinfo
  {pages} {211--217} (\bibinfo {year} {2020})}\BibitemShut {NoStop}%
\bibitem [{\citenamefont {Albiez}\ \emph {et~al.}(2005)\citenamefont {Albiez},
  \citenamefont {Gati}, \citenamefont {F{\"o}lling}, \citenamefont {Hunsmann},
  \citenamefont {Cristiani},\ and\ \citenamefont
  {Oberthaler}}]{albiez2005direct_jjtunneling}%
  \BibitemOpen
  \bibfield  {author} {\bibinfo {author} {\bibfnamefont {M.}~\bibnamefont
  {Albiez}}, \bibinfo {author} {\bibfnamefont {R.}~\bibnamefont {Gati}},
  \bibinfo {author} {\bibfnamefont {J.}~\bibnamefont {F{\"o}lling}}, \bibinfo
  {author} {\bibfnamefont {S.}~\bibnamefont {Hunsmann}}, \bibinfo {author}
  {\bibfnamefont {M.}~\bibnamefont {Cristiani}}, \ and\ \bibinfo {author}
  {\bibfnamefont {M.~K.}\ \bibnamefont {Oberthaler}},\ }\bibfield  {title}
  {\enquote {\bibinfo {title} {Direct observation of tunneling and nonlinear
  self-trapping in a single bosonic josephson junction},}\ }\href@noop {}
  {\bibfield  {journal} {\bibinfo  {journal} {Physical review letters}\
  }\textbf {\bibinfo {volume} {95}},\ \bibinfo {pages} {010402} (\bibinfo
  {year} {2005})}\BibitemShut {NoStop}%
\bibitem [{\citenamefont {Zhou}\ \emph {et~al.}(2014)\citenamefont {Zhou},
  \citenamefont {Schmitt}, \citenamefont {Bertet}, \citenamefont {Vion},
  \citenamefont {Wustmann}, \citenamefont {Shumeiko},\ and\ \citenamefont
  {Est{\`e}ve}}]{zhou2014high_nonlinearJPA}%
  \BibitemOpen
  \bibfield  {author} {\bibinfo {author} {\bibfnamefont {X.}~\bibnamefont
  {Zhou}}, \bibinfo {author} {\bibfnamefont {V.}~\bibnamefont {Schmitt}},
  \bibinfo {author} {\bibfnamefont {P.}~\bibnamefont {Bertet}}, \bibinfo
  {author} {\bibfnamefont {D.}~\bibnamefont {Vion}}, \bibinfo {author}
  {\bibfnamefont {W.}~\bibnamefont {Wustmann}}, \bibinfo {author}
  {\bibfnamefont {V.}~\bibnamefont {Shumeiko}}, \ and\ \bibinfo {author}
  {\bibfnamefont {D.}~\bibnamefont {Est{\`e}ve}},\ }\bibfield  {title}
  {\enquote {\bibinfo {title} {High-gain weakly nonlinear flux-modulated
  josephson parametric amplifier using a squid array},}\ }\href@noop {}
  {\bibfield  {journal} {\bibinfo  {journal} {Physical Review B}\ }\textbf
  {\bibinfo {volume} {89}},\ \bibinfo {pages} {214517} (\bibinfo {year}
  {2014})}\BibitemShut {NoStop}%
\bibitem [{\citenamefont {Castellanos-Beltran}\ \emph
  {et~al.}(2008)\citenamefont {Castellanos-Beltran}, \citenamefont {Irwin},
  \citenamefont {Hilton}, \citenamefont {Vale},\ and\ \citenamefont
  {Lehnert}}]{castellanos2008amplification}%
  \BibitemOpen
  \bibfield  {author} {\bibinfo {author} {\bibfnamefont {M.}~\bibnamefont
  {Castellanos-Beltran}}, \bibinfo {author} {\bibfnamefont {K.}~\bibnamefont
  {Irwin}}, \bibinfo {author} {\bibfnamefont {G.}~\bibnamefont {Hilton}},
  \bibinfo {author} {\bibfnamefont {L.}~\bibnamefont {Vale}}, \ and\ \bibinfo
  {author} {\bibfnamefont {K.}~\bibnamefont {Lehnert}},\ }\bibfield  {title}
  {\enquote {\bibinfo {title} {Amplification and squeezing of quantum noise
  with a tunable josephson metamaterial},}\ }\href@noop {} {\bibfield
  {journal} {\bibinfo  {journal} {Nature Physics}\ }\textbf {\bibinfo {volume}
  {4}},\ \bibinfo {pages} {929--931} (\bibinfo {year} {2008})}\BibitemShut
  {NoStop}%
\bibitem [{\citenamefont {Abdo}\ \emph {et~al.}(2013)\citenamefont {Abdo},
  \citenamefont {Sliwa}, \citenamefont {Schackert}, \citenamefont {Bergeal},
  \citenamefont {Hatridge}, \citenamefont {Frunzio}, \citenamefont {Stone},\
  and\ \citenamefont {Devoret}}]{abdo2013full_freqconv}%
  \BibitemOpen
  \bibfield  {author} {\bibinfo {author} {\bibfnamefont {B.}~\bibnamefont
  {Abdo}}, \bibinfo {author} {\bibfnamefont {K.}~\bibnamefont {Sliwa}},
  \bibinfo {author} {\bibfnamefont {F.}~\bibnamefont {Schackert}}, \bibinfo
  {author} {\bibfnamefont {N.}~\bibnamefont {Bergeal}}, \bibinfo {author}
  {\bibfnamefont {M.}~\bibnamefont {Hatridge}}, \bibinfo {author}
  {\bibfnamefont {L.}~\bibnamefont {Frunzio}}, \bibinfo {author} {\bibfnamefont
  {A.~D.}\ \bibnamefont {Stone}}, \ and\ \bibinfo {author} {\bibfnamefont
  {M.}~\bibnamefont {Devoret}},\ }\bibfield  {title} {\enquote {\bibinfo
  {title} {Full coherent frequency conversion between two propagating microwave
  modes},}\ }\href@noop {} {\bibfield  {journal} {\bibinfo  {journal} {Physical
  review letters}\ }\textbf {\bibinfo {volume} {110}},\ \bibinfo {pages}
  {173902} (\bibinfo {year} {2013})}\BibitemShut {NoStop}%
\bibitem [{\citenamefont {Zakka-Bajjani}\ \emph {et~al.}(2011)\citenamefont
  {Zakka-Bajjani}, \citenamefont {Nguyen}, \citenamefont {Lee}, \citenamefont
  {Vale}, \citenamefont {Simmonds},\ and\ \citenamefont
  {Aumentado}}]{zakka2011quantum_freqconv}%
  \BibitemOpen
  \bibfield  {author} {\bibinfo {author} {\bibfnamefont {E.}~\bibnamefont
  {Zakka-Bajjani}}, \bibinfo {author} {\bibfnamefont {F.}~\bibnamefont
  {Nguyen}}, \bibinfo {author} {\bibfnamefont {M.}~\bibnamefont {Lee}},
  \bibinfo {author} {\bibfnamefont {L.~R.}\ \bibnamefont {Vale}}, \bibinfo
  {author} {\bibfnamefont {R.~W.}\ \bibnamefont {Simmonds}}, \ and\ \bibinfo
  {author} {\bibfnamefont {J.}~\bibnamefont {Aumentado}},\ }\bibfield  {title}
  {\enquote {\bibinfo {title} {Quantum superposition of a single microwave
  photon in two different’colour’states},}\ }\href@noop {} {\bibfield
  {journal} {\bibinfo  {journal} {Nature Physics}\ }\textbf {\bibinfo {volume}
  {7}},\ \bibinfo {pages} {599--603} (\bibinfo {year} {2011})}\BibitemShut
  {NoStop}%
\bibitem [{\citenamefont {Sirois}\ \emph {et~al.}(2015)\citenamefont {Sirois},
  \citenamefont {Castellanos-Beltran}, \citenamefont {DeFeo}, \citenamefont
  {Ranzani}, \citenamefont {Lecocq}, \citenamefont {Simmonds}, \citenamefont
  {Teufel},\ and\ \citenamefont {Aumentado}}]{sirois2015coherent_freqconv}%
  \BibitemOpen
  \bibfield  {author} {\bibinfo {author} {\bibfnamefont {A.~J.}\ \bibnamefont
  {Sirois}}, \bibinfo {author} {\bibfnamefont {M.}~\bibnamefont
  {Castellanos-Beltran}}, \bibinfo {author} {\bibfnamefont {M.}~\bibnamefont
  {DeFeo}}, \bibinfo {author} {\bibfnamefont {L.}~\bibnamefont {Ranzani}},
  \bibinfo {author} {\bibfnamefont {F.}~\bibnamefont {Lecocq}}, \bibinfo
  {author} {\bibfnamefont {R.}~\bibnamefont {Simmonds}}, \bibinfo {author}
  {\bibfnamefont {J.}~\bibnamefont {Teufel}}, \ and\ \bibinfo {author}
  {\bibfnamefont {J.}~\bibnamefont {Aumentado}},\ }\bibfield  {title} {\enquote
  {\bibinfo {title} {Coherent-state storage and retrieval between
  superconducting cavities using parametric frequency conversion},}\
  }\href@noop {} {\bibfield  {journal} {\bibinfo  {journal} {Applied Physics
  Letters}\ }\textbf {\bibinfo {volume} {106}},\ \bibinfo {pages} {172603}
  (\bibinfo {year} {2015})}\BibitemShut {NoStop}%
\bibitem [{\citenamefont {Svensson}\ \emph {et~al.}(2018)\citenamefont
  {Svensson}, \citenamefont {Bengtsson}, \citenamefont {Bylander},
  \citenamefont {Shumeiko},\ and\ \citenamefont
  {Delsing}}]{svensson2018period_subharmonic}%
  \BibitemOpen
  \bibfield  {author} {\bibinfo {author} {\bibfnamefont {I.-M.}\ \bibnamefont
  {Svensson}}, \bibinfo {author} {\bibfnamefont {A.}~\bibnamefont {Bengtsson}},
  \bibinfo {author} {\bibfnamefont {J.}~\bibnamefont {Bylander}}, \bibinfo
  {author} {\bibfnamefont {V.}~\bibnamefont {Shumeiko}}, \ and\ \bibinfo
  {author} {\bibfnamefont {P.}~\bibnamefont {Delsing}},\ }\bibfield  {title}
  {\enquote {\bibinfo {title} {Period multiplication in a parametrically driven
  superconducting resonator},}\ }\href@noop {} {\bibfield  {journal} {\bibinfo
  {journal} {Applied Physics Letters}\ }\textbf {\bibinfo {volume} {113}},\
  \bibinfo {pages} {022602} (\bibinfo {year} {2018})}\BibitemShut {NoStop}%
\bibitem [{\citenamefont {Svensson}\ \emph {et~al.}(2017)\citenamefont
  {Svensson}, \citenamefont {Bengtsson}, \citenamefont {Krantz}, \citenamefont
  {Bylander}, \citenamefont {Shumeiko},\ and\ \citenamefont
  {Delsing}}]{svensson2017period_harmonic}%
  \BibitemOpen
  \bibfield  {author} {\bibinfo {author} {\bibfnamefont {I.-M.}\ \bibnamefont
  {Svensson}}, \bibinfo {author} {\bibfnamefont {A.}~\bibnamefont {Bengtsson}},
  \bibinfo {author} {\bibfnamefont {P.}~\bibnamefont {Krantz}}, \bibinfo
  {author} {\bibfnamefont {J.}~\bibnamefont {Bylander}}, \bibinfo {author}
  {\bibfnamefont {V.}~\bibnamefont {Shumeiko}}, \ and\ \bibinfo {author}
  {\bibfnamefont {P.}~\bibnamefont {Delsing}},\ }\bibfield  {title} {\enquote
  {\bibinfo {title} {Period-tripling subharmonic oscillations in a driven
  superconducting resonator},}\ }\href@noop {} {\bibfield  {journal} {\bibinfo
  {journal} {Physical Review B}\ }\textbf {\bibinfo {volume} {96}},\ \bibinfo
  {pages} {174503} (\bibinfo {year} {2017})}\BibitemShut {NoStop}%
\bibitem [{\citenamefont {Vlastakis}\ \emph {et~al.}(2013)\citenamefont
  {Vlastakis}, \citenamefont {Kirchmair}, \citenamefont {Leghtas},
  \citenamefont {Nigg}, \citenamefont {Frunzio}, \citenamefont {Girvin},
  \citenamefont {Mirrahimi}, \citenamefont {Devoret},\ and\ \citenamefont
  {Schoelkopf}}]{vlastakis2013multiphoton}%
  \BibitemOpen
  \bibfield  {author} {\bibinfo {author} {\bibfnamefont {B.}~\bibnamefont
  {Vlastakis}}, \bibinfo {author} {\bibfnamefont {G.}~\bibnamefont
  {Kirchmair}}, \bibinfo {author} {\bibfnamefont {Z.}~\bibnamefont {Leghtas}},
  \bibinfo {author} {\bibfnamefont {S.~E.}\ \bibnamefont {Nigg}}, \bibinfo
  {author} {\bibfnamefont {L.}~\bibnamefont {Frunzio}}, \bibinfo {author}
  {\bibfnamefont {S.~M.}\ \bibnamefont {Girvin}}, \bibinfo {author}
  {\bibfnamefont {M.}~\bibnamefont {Mirrahimi}}, \bibinfo {author}
  {\bibfnamefont {M.~H.}\ \bibnamefont {Devoret}}, \ and\ \bibinfo {author}
  {\bibfnamefont {R.~J.}\ \bibnamefont {Schoelkopf}},\ }\bibfield  {title}
  {\enquote {\bibinfo {title} {Deterministically encoding quantum information
  using 100-photon schr{\"o}dinger cat states},}\ }\href@noop {} {\bibfield
  {journal} {\bibinfo  {journal} {Science}\ }\textbf {\bibinfo {volume}
  {342}},\ \bibinfo {pages} {607--610} (\bibinfo {year} {2013})}\BibitemShut
  {NoStop}%
\bibitem [{\citenamefont {Leghtas}\ \emph {et~al.}(2015)\citenamefont
  {Leghtas}, \citenamefont {Touzard}, \citenamefont {Pop}, \citenamefont {Kou},
  \citenamefont {Vlastakis}, \citenamefont {Petrenko}, \citenamefont {Sliwa},
  \citenamefont {Narla}, \citenamefont {Shankar}, \citenamefont {Hatridge}
  \emph {et~al.}}]{leghtas2015confining_multi}%
  \BibitemOpen
  \bibfield  {author} {\bibinfo {author} {\bibfnamefont {Z.}~\bibnamefont
  {Leghtas}}, \bibinfo {author} {\bibfnamefont {S.}~\bibnamefont {Touzard}},
  \bibinfo {author} {\bibfnamefont {I.~M.}\ \bibnamefont {Pop}}, \bibinfo
  {author} {\bibfnamefont {A.}~\bibnamefont {Kou}}, \bibinfo {author}
  {\bibfnamefont {B.}~\bibnamefont {Vlastakis}}, \bibinfo {author}
  {\bibfnamefont {A.}~\bibnamefont {Petrenko}}, \bibinfo {author}
  {\bibfnamefont {K.~M.}\ \bibnamefont {Sliwa}}, \bibinfo {author}
  {\bibfnamefont {A.}~\bibnamefont {Narla}}, \bibinfo {author} {\bibfnamefont
  {S.}~\bibnamefont {Shankar}}, \bibinfo {author} {\bibfnamefont {M.~J.}\
  \bibnamefont {Hatridge}},  \emph {et~al.},\ }\bibfield  {title} {\enquote
  {\bibinfo {title} {Confining the state of light to a quantum manifold by
  engineered two-photon loss},}\ }\href@noop {} {\bibfield  {journal} {\bibinfo
   {journal} {Science}\ }\textbf {\bibinfo {volume} {347}},\ \bibinfo {pages}
  {853--857} (\bibinfo {year} {2015})}\BibitemShut {NoStop}%
\bibitem [{\citenamefont {Kirchmair}\ \emph {et~al.}(2013)\citenamefont
  {Kirchmair}, \citenamefont {Vlastakis}, \citenamefont {Leghtas},
  \citenamefont {Nigg}, \citenamefont {Paik}, \citenamefont {Ginossar},
  \citenamefont {Mirrahimi}, \citenamefont {Frunzio}, \citenamefont {Girvin},\
  and\ \citenamefont {Schoelkopf}}]{kirchmair2013observation_kerr}%
  \BibitemOpen
  \bibfield  {author} {\bibinfo {author} {\bibfnamefont {G.}~\bibnamefont
  {Kirchmair}}, \bibinfo {author} {\bibfnamefont {B.}~\bibnamefont
  {Vlastakis}}, \bibinfo {author} {\bibfnamefont {Z.}~\bibnamefont {Leghtas}},
  \bibinfo {author} {\bibfnamefont {S.~E.}\ \bibnamefont {Nigg}}, \bibinfo
  {author} {\bibfnamefont {H.}~\bibnamefont {Paik}}, \bibinfo {author}
  {\bibfnamefont {E.}~\bibnamefont {Ginossar}}, \bibinfo {author}
  {\bibfnamefont {M.}~\bibnamefont {Mirrahimi}}, \bibinfo {author}
  {\bibfnamefont {L.}~\bibnamefont {Frunzio}}, \bibinfo {author} {\bibfnamefont
  {S.~M.}\ \bibnamefont {Girvin}}, \ and\ \bibinfo {author} {\bibfnamefont
  {R.~J.}\ \bibnamefont {Schoelkopf}},\ }\bibfield  {title} {\enquote {\bibinfo
  {title} {Observation of quantum state collapse and revival due to the
  single-photon kerr effect},}\ }\href@noop {} {\bibfield  {journal} {\bibinfo
  {journal} {Nature}\ }\textbf {\bibinfo {volume} {495}},\ \bibinfo {pages}
  {205--209} (\bibinfo {year} {2013})}\BibitemShut {NoStop}%
\bibitem [{\citenamefont {Yin}\ \emph {et~al.}(2012)\citenamefont {Yin},
  \citenamefont {Wang}, \citenamefont {Mariantoni}, \citenamefont {Bialczak},
  \citenamefont {Barends}, \citenamefont {Chen}, \citenamefont {Lenander},
  \citenamefont {Lucero}, \citenamefont {Neeley}, \citenamefont {O'Connell}
  \emph {et~al.}}]{yin2012dynamic_Kerr}%
  \BibitemOpen
  \bibfield  {author} {\bibinfo {author} {\bibfnamefont {Y.}~\bibnamefont
  {Yin}}, \bibinfo {author} {\bibfnamefont {H.}~\bibnamefont {Wang}}, \bibinfo
  {author} {\bibfnamefont {M.}~\bibnamefont {Mariantoni}}, \bibinfo {author}
  {\bibfnamefont {R.~C.}\ \bibnamefont {Bialczak}}, \bibinfo {author}
  {\bibfnamefont {R.}~\bibnamefont {Barends}}, \bibinfo {author} {\bibfnamefont
  {Y.}~\bibnamefont {Chen}}, \bibinfo {author} {\bibfnamefont {M.}~\bibnamefont
  {Lenander}}, \bibinfo {author} {\bibfnamefont {E.}~\bibnamefont {Lucero}},
  \bibinfo {author} {\bibfnamefont {M.}~\bibnamefont {Neeley}}, \bibinfo
  {author} {\bibfnamefont {A.}~\bibnamefont {O'Connell}},  \emph {et~al.},\
  }\bibfield  {title} {\enquote {\bibinfo {title} {Dynamic quantum kerr effect
  in circuit quantum electrodynamics},}\ }\href@noop {} {\bibfield  {journal}
  {\bibinfo  {journal} {Physical Review A}\ }\textbf {\bibinfo {volume} {85}},\
  \bibinfo {pages} {023826} (\bibinfo {year} {2012})}\BibitemShut {NoStop}%
\bibitem [{\citenamefont {Goto}(2016)}]{goto2016bifurcation_goto}%
  \BibitemOpen
  \bibfield  {author} {\bibinfo {author} {\bibfnamefont {H.}~\bibnamefont
  {Goto}},\ }\bibfield  {title} {\enquote {\bibinfo {title} {Bifurcation-based
  adiabatic quantum computation with a nonlinear oscillator network},}\
  }\href@noop {} {\bibfield  {journal} {\bibinfo  {journal} {Scientific
  reports}\ }\textbf {\bibinfo {volume} {6}},\ \bibinfo {pages} {1--8}
  (\bibinfo {year} {2016})}\BibitemShut {NoStop}%
\bibitem [{\citenamefont {Goto}(2019)}]{goto2019quantum}%
  \BibitemOpen
  \bibfield  {author} {\bibinfo {author} {\bibfnamefont {H.}~\bibnamefont
  {Goto}},\ }\bibfield  {title} {\enquote {\bibinfo {title} {Quantum
  computation based on quantum adiabatic bifurcations of kerr-nonlinear
  parametric oscillators},}\ }\href@noop {} {\bibfield  {journal} {\bibinfo
  {journal} {Journal of the Physical Society of Japan}\ }\textbf {\bibinfo
  {volume} {88}},\ \bibinfo {pages} {061015} (\bibinfo {year}
  {2019})}\BibitemShut {NoStop}%
\bibitem [{\citenamefont {Wang}\ \emph {et~al.}(2019)\citenamefont {Wang},
  \citenamefont {Pechal}, \citenamefont {Wollack}, \citenamefont
  {Arrangoiz-Arriola}, \citenamefont {Gao}, \citenamefont {Lee},\ and\
  \citenamefont {Safavi-Naeini}}]{wang2019quantum_patricio}%
  \BibitemOpen
  \bibfield  {author} {\bibinfo {author} {\bibfnamefont {Z.}~\bibnamefont
  {Wang}}, \bibinfo {author} {\bibfnamefont {M.}~\bibnamefont {Pechal}},
  \bibinfo {author} {\bibfnamefont {E.~A.}\ \bibnamefont {Wollack}}, \bibinfo
  {author} {\bibfnamefont {P.}~\bibnamefont {Arrangoiz-Arriola}}, \bibinfo
  {author} {\bibfnamefont {M.}~\bibnamefont {Gao}}, \bibinfo {author}
  {\bibfnamefont {N.~R.}\ \bibnamefont {Lee}}, \ and\ \bibinfo {author}
  {\bibfnamefont {A.~H.}\ \bibnamefont {Safavi-Naeini}},\ }\bibfield  {title}
  {\enquote {\bibinfo {title} {Quantum dynamics of a few-photon parametric
  oscillator},}\ }\href@noop {} {\bibfield  {journal} {\bibinfo  {journal}
  {Physical Review X}\ }\textbf {\bibinfo {volume} {9}},\ \bibinfo {pages}
  {021049} (\bibinfo {year} {2019})}\BibitemShut {NoStop}%
\bibitem [{\citenamefont {Grimm}\ \emph {et~al.}(2020)\citenamefont {Grimm},
  \citenamefont {Frattini}, \citenamefont {Puri}, \citenamefont {Mundhada},
  \citenamefont {Touzard}, \citenamefont {Mirrahimi}, \citenamefont {Girvin},
  \citenamefont {Shankar},\ and\ \citenamefont
  {Devoret}}]{grimm2020stabilization_puri_bifur}%
  \BibitemOpen
  \bibfield  {author} {\bibinfo {author} {\bibfnamefont {A.}~\bibnamefont
  {Grimm}}, \bibinfo {author} {\bibfnamefont {N.~E.}\ \bibnamefont {Frattini}},
  \bibinfo {author} {\bibfnamefont {S.}~\bibnamefont {Puri}}, \bibinfo {author}
  {\bibfnamefont {S.~O.}\ \bibnamefont {Mundhada}}, \bibinfo {author}
  {\bibfnamefont {S.}~\bibnamefont {Touzard}}, \bibinfo {author} {\bibfnamefont
  {M.}~\bibnamefont {Mirrahimi}}, \bibinfo {author} {\bibfnamefont {S.~M.}\
  \bibnamefont {Girvin}}, \bibinfo {author} {\bibfnamefont {S.}~\bibnamefont
  {Shankar}}, \ and\ \bibinfo {author} {\bibfnamefont {M.~H.}\ \bibnamefont
  {Devoret}},\ }\bibfield  {title} {\enquote {\bibinfo {title} {Stabilization
  and operation of a kerr-cat qubit},}\ }\href@noop {} {\bibfield  {journal}
  {\bibinfo  {journal} {Nature}\ }\textbf {\bibinfo {volume} {584}},\ \bibinfo
  {pages} {205--209} (\bibinfo {year} {2020})}\BibitemShut {NoStop}%
\bibitem [{\citenamefont {Frattini}\ \emph {et~al.}(2022)\citenamefont
  {Frattini}, \citenamefont {Corti{\~n}as}, \citenamefont {Venkatraman},
  \citenamefont {Xiao}, \citenamefont {Su}, \citenamefont {Lei}, \citenamefont
  {Chapman}, \citenamefont {Joshi}, \citenamefont {Girvin}, \citenamefont
  {Schoelkopf} \emph {et~al.}}]{frattini2022squeezed}%
  \BibitemOpen
  \bibfield  {author} {\bibinfo {author} {\bibfnamefont {N.~E.}\ \bibnamefont
  {Frattini}}, \bibinfo {author} {\bibfnamefont {R.~G.}\ \bibnamefont
  {Corti{\~n}as}}, \bibinfo {author} {\bibfnamefont {J.}~\bibnamefont
  {Venkatraman}}, \bibinfo {author} {\bibfnamefont {X.}~\bibnamefont {Xiao}},
  \bibinfo {author} {\bibfnamefont {Q.}~\bibnamefont {Su}}, \bibinfo {author}
  {\bibfnamefont {C.~U.}\ \bibnamefont {Lei}}, \bibinfo {author} {\bibfnamefont
  {B.~J.}\ \bibnamefont {Chapman}}, \bibinfo {author} {\bibfnamefont {V.~R.}\
  \bibnamefont {Joshi}}, \bibinfo {author} {\bibfnamefont {S.}~\bibnamefont
  {Girvin}}, \bibinfo {author} {\bibfnamefont {R.~J.}\ \bibnamefont
  {Schoelkopf}},  \emph {et~al.},\ }\bibfield  {title} {\enquote {\bibinfo
  {title} {The squeezed kerr oscillator: spectral kissing and phase-flip
  robustness},}\ }\href@noop {} {\bibfield  {journal} {\bibinfo  {journal}
  {arXiv preprint arXiv:2209.03934}\ } (\bibinfo {year} {2022})}\BibitemShut
  {NoStop}%
\bibitem [{\citenamefont {Reid}\ and\ \citenamefont
  {Yurke}(1992)}]{reid1992effect_yurke}%
  \BibitemOpen
  \bibfield  {author} {\bibinfo {author} {\bibfnamefont {M.}~\bibnamefont
  {Reid}}\ and\ \bibinfo {author} {\bibfnamefont {B.}~\bibnamefont {Yurke}},\
  }\bibfield  {title} {\enquote {\bibinfo {title} {Effect of bistability and
  superpositions on quantum statistics in degenerate parametric oscillation},}\
  }\href@noop {} {\bibfield  {journal} {\bibinfo  {journal} {Physical Review
  A}\ }\textbf {\bibinfo {volume} {46}},\ \bibinfo {pages} {4131} (\bibinfo
  {year} {1992})}\BibitemShut {NoStop}%
\bibitem [{\citenamefont {Yamamoto}, \citenamefont {Koshino},\ and\
  \citenamefont {Nakamura}(2016)}]{yamamoto2016parametric}%
  \BibitemOpen
  \bibfield  {author} {\bibinfo {author} {\bibfnamefont {T.}~\bibnamefont
  {Yamamoto}}, \bibinfo {author} {\bibfnamefont {K.}~\bibnamefont {Koshino}}, \
  and\ \bibinfo {author} {\bibfnamefont {Y.}~\bibnamefont {Nakamura}},\
  }\bibfield  {title} {\enquote {\bibinfo {title} {Parametric amplifier and
  oscillator based on josephson junction circuitry},}\ }in\ \href@noop {}
  {\emph {\bibinfo {booktitle} {Principles and Methods of Quantum Information
  Technologies}}}\ (\bibinfo  {publisher} {Springer},\ \bibinfo {year} {2016})\
  pp.\ \bibinfo {pages} {495--513}\BibitemShut {NoStop}%
\bibitem [{\citenamefont {Lin}\ \emph {et~al.}(2014)\citenamefont {Lin},
  \citenamefont {Inomata}, \citenamefont {Koshino}, \citenamefont {Oliver},
  \citenamefont {Nakamura}, \citenamefont {Tsai},\ and\ \citenamefont
  {Yamamoto}}]{lin2014josephson}%
  \BibitemOpen
  \bibfield  {author} {\bibinfo {author} {\bibfnamefont {Z.}~\bibnamefont
  {Lin}}, \bibinfo {author} {\bibfnamefont {K.}~\bibnamefont {Inomata}},
  \bibinfo {author} {\bibfnamefont {K.}~\bibnamefont {Koshino}}, \bibinfo
  {author} {\bibfnamefont {W.}~\bibnamefont {Oliver}}, \bibinfo {author}
  {\bibfnamefont {Y.}~\bibnamefont {Nakamura}}, \bibinfo {author}
  {\bibfnamefont {J.-S.}\ \bibnamefont {Tsai}}, \ and\ \bibinfo {author}
  {\bibfnamefont {T.}~\bibnamefont {Yamamoto}},\ }\bibfield  {title} {\enquote
  {\bibinfo {title} {Josephson parametric phase-locked oscillator and its
  application to dispersive readout of superconducting qubits},}\ }\href@noop
  {} {\bibfield  {journal} {\bibinfo  {journal} {Nature communications}\
  }\textbf {\bibinfo {volume} {5}},\ \bibinfo {pages} {1--6} (\bibinfo {year}
  {2014})}\BibitemShut {NoStop}%
\bibitem [{\citenamefont {Dykman}(2012)}]{dykman2012fluctuating}%
  \BibitemOpen
  \bibfield  {author} {\bibinfo {author} {\bibfnamefont {M.}~\bibnamefont
  {Dykman}},\ }\href@noop {} {\emph {\bibinfo {title} {Fluctuating nonlinear
  oscillators: from nanomechanics to quantum superconducting circuits}}}\
  (\bibinfo  {publisher} {Oxford University Press},\ \bibinfo {year}
  {2012})\BibitemShut {NoStop}%
\bibitem [{\citenamefont {Wilson}\ \emph {et~al.}(2010)\citenamefont {Wilson},
  \citenamefont {Duty}, \citenamefont {Sandberg}, \citenamefont {Persson},
  \citenamefont {Shumeiko},\ and\ \citenamefont
  {Delsing}}]{wilson2010photon_dynamical}%
  \BibitemOpen
  \bibfield  {author} {\bibinfo {author} {\bibfnamefont {C.}~\bibnamefont
  {Wilson}}, \bibinfo {author} {\bibfnamefont {T.}~\bibnamefont {Duty}},
  \bibinfo {author} {\bibfnamefont {M.}~\bibnamefont {Sandberg}}, \bibinfo
  {author} {\bibfnamefont {F.}~\bibnamefont {Persson}}, \bibinfo {author}
  {\bibfnamefont {V.}~\bibnamefont {Shumeiko}}, \ and\ \bibinfo {author}
  {\bibfnamefont {P.}~\bibnamefont {Delsing}},\ }\bibfield  {title} {\enquote
  {\bibinfo {title} {Photon generation in an electromagnetic cavity with a
  time-dependent boundary},}\ }\href@noop {} {\bibfield  {journal} {\bibinfo
  {journal} {Physical review letters}\ }\textbf {\bibinfo {volume} {105}},\
  \bibinfo {pages} {233907} (\bibinfo {year} {2010})}\BibitemShut {NoStop}%
\bibitem [{\citenamefont {Wolinsky}\ and\ \citenamefont
  {Carmichael}(1988)}]{wolinsky1988quantum_noise_JPO}%
  \BibitemOpen
  \bibfield  {author} {\bibinfo {author} {\bibfnamefont {M.}~\bibnamefont
  {Wolinsky}}\ and\ \bibinfo {author} {\bibfnamefont {H.}~\bibnamefont
  {Carmichael}},\ }\bibfield  {title} {\enquote {\bibinfo {title} {Quantum
  noise in the parametric oscillator: from squeezed states to coherent-state
  superpositions},}\ }\href@noop {} {\bibfield  {journal} {\bibinfo  {journal}
  {Physical review letters}\ }\textbf {\bibinfo {volume} {60}},\ \bibinfo
  {pages} {1836} (\bibinfo {year} {1988})}\BibitemShut {NoStop}%
\bibitem [{\citenamefont {Kinsler}\ and\ \citenamefont
  {Drummond}(1991)}]{kinsler1991quantum_switching}%
  \BibitemOpen
  \bibfield  {author} {\bibinfo {author} {\bibfnamefont {P.}~\bibnamefont
  {Kinsler}}\ and\ \bibinfo {author} {\bibfnamefont {P.~D.}\ \bibnamefont
  {Drummond}},\ }\bibfield  {title} {\enquote {\bibinfo {title} {Quantum
  dynamics of the parametric oscillator},}\ }\href@noop {} {\bibfield
  {journal} {\bibinfo  {journal} {Physical Review A}\ }\textbf {\bibinfo
  {volume} {43}},\ \bibinfo {pages} {6194} (\bibinfo {year}
  {1991})}\BibitemShut {NoStop}%
\bibitem [{\citenamefont {Bhai}\ \emph {et~al.}(2022)\citenamefont {Bhai},
  \citenamefont {Mukai}, \citenamefont {Yamamoto},\ and\ \citenamefont
  {Tsai}}]{bhai2022noise}%
  \BibitemOpen
  \bibfield  {author} {\bibinfo {author} {\bibfnamefont {G.~L.}\ \bibnamefont
  {Bhai}}, \bibinfo {author} {\bibfnamefont {H.}~\bibnamefont {Mukai}},
  \bibinfo {author} {\bibfnamefont {T.}~\bibnamefont {Yamamoto}}, \ and\
  \bibinfo {author} {\bibfnamefont {J.-S.}\ \bibnamefont {Tsai}},\ }\bibfield
  {title} {\enquote {\bibinfo {title} {Noise properties of a josephson
  parametric oscillator},}\ }\href@noop {} {\bibfield  {journal} {\bibinfo
  {journal} {arXiv:2210.15116}\ } (\bibinfo {year} {2022})}\BibitemShut
  {NoStop}%
\bibitem [{\citenamefont {Krantz}\ \emph {et~al.}(2016)\citenamefont {Krantz},
  \citenamefont {Bengtsson}, \citenamefont {Simoen}, \citenamefont
  {Gustavsson}, \citenamefont {Shumeiko}, \citenamefont {Oliver}, \citenamefont
  {Wilson}, \citenamefont {Delsing},\ and\ \citenamefont
  {Bylander}}]{krantz2016single}%
  \BibitemOpen
  \bibfield  {author} {\bibinfo {author} {\bibfnamefont {P.}~\bibnamefont
  {Krantz}}, \bibinfo {author} {\bibfnamefont {A.}~\bibnamefont {Bengtsson}},
  \bibinfo {author} {\bibfnamefont {M.}~\bibnamefont {Simoen}}, \bibinfo
  {author} {\bibfnamefont {S.}~\bibnamefont {Gustavsson}}, \bibinfo {author}
  {\bibfnamefont {V.}~\bibnamefont {Shumeiko}}, \bibinfo {author}
  {\bibfnamefont {W.}~\bibnamefont {Oliver}}, \bibinfo {author} {\bibfnamefont
  {C.}~\bibnamefont {Wilson}}, \bibinfo {author} {\bibfnamefont
  {P.}~\bibnamefont {Delsing}}, \ and\ \bibinfo {author} {\bibfnamefont
  {J.}~\bibnamefont {Bylander}},\ }\bibfield  {title} {\enquote {\bibinfo
  {title} {Single-shot read-out of a superconducting qubit using a josephson
  parametric oscillator},}\ }\href@noop {} {\bibfield  {journal} {\bibinfo
  {journal} {Nature communications}\ }\textbf {\bibinfo {volume} {7}},\
  \bibinfo {pages} {1--8} (\bibinfo {year} {2016})}\BibitemShut {NoStop}%
\bibitem [{\citenamefont {Bengtsson}\ \emph {et~al.}(2018)\citenamefont
  {Bengtsson}, \citenamefont {Krantz}, \citenamefont {Simoen}, \citenamefont
  {Svensson}, \citenamefont {Schneider}, \citenamefont {Shumeiko},
  \citenamefont {Delsing},\ and\ \citenamefont
  {Bylander}}]{bengtsson2018nondegenerate}%
  \BibitemOpen
  \bibfield  {author} {\bibinfo {author} {\bibfnamefont {A.}~\bibnamefont
  {Bengtsson}}, \bibinfo {author} {\bibfnamefont {P.}~\bibnamefont {Krantz}},
  \bibinfo {author} {\bibfnamefont {M.}~\bibnamefont {Simoen}}, \bibinfo
  {author} {\bibfnamefont {I.-M.}\ \bibnamefont {Svensson}}, \bibinfo {author}
  {\bibfnamefont {B.}~\bibnamefont {Schneider}}, \bibinfo {author}
  {\bibfnamefont {V.}~\bibnamefont {Shumeiko}}, \bibinfo {author}
  {\bibfnamefont {P.}~\bibnamefont {Delsing}}, \ and\ \bibinfo {author}
  {\bibfnamefont {J.}~\bibnamefont {Bylander}},\ }\bibfield  {title} {\enquote
  {\bibinfo {title} {Nondegenerate parametric oscillations in a tunable
  superconducting resonator},}\ }\href@noop {} {\bibfield  {journal} {\bibinfo
  {journal} {Physical Review B}\ }\textbf {\bibinfo {volume} {97}},\ \bibinfo
  {pages} {144502} (\bibinfo {year} {2018})}\BibitemShut {NoStop}%
\bibitem [{\citenamefont {Danner}\ \emph {et~al.}(2021)\citenamefont {Danner},
  \citenamefont {Padurariu}, \citenamefont {Ankerhold},\ and\ \citenamefont
  {Kubala}}]{danner2021injection_photonic}%
  \BibitemOpen
  \bibfield  {author} {\bibinfo {author} {\bibfnamefont {L.}~\bibnamefont
  {Danner}}, \bibinfo {author} {\bibfnamefont {C.}~\bibnamefont {Padurariu}},
  \bibinfo {author} {\bibfnamefont {J.}~\bibnamefont {Ankerhold}}, \ and\
  \bibinfo {author} {\bibfnamefont {B.}~\bibnamefont {Kubala}},\ }\bibfield
  {title} {\enquote {\bibinfo {title} {Injection locking and synchronization in
  josephson photonics devices},}\ }\href@noop {} {\bibfield  {journal}
  {\bibinfo  {journal} {Physical Review B}\ }\textbf {\bibinfo {volume}
  {104}},\ \bibinfo {pages} {054517} (\bibinfo {year} {2021})}\BibitemShut
  {NoStop}%
\bibitem [{\citenamefont {T{\'o}th}\ \emph {et~al.}(2018)\citenamefont
  {T{\'o}th}, \citenamefont {Bernier}, \citenamefont {Feofanov},\ and\
  \citenamefont {Kippenberg}}]{toth2018maser}%
  \BibitemOpen
  \bibfield  {author} {\bibinfo {author} {\bibfnamefont {L.}~\bibnamefont
  {T{\'o}th}}, \bibinfo {author} {\bibfnamefont {N.}~\bibnamefont {Bernier}},
  \bibinfo {author} {\bibfnamefont {A.}~\bibnamefont {Feofanov}}, \ and\
  \bibinfo {author} {\bibfnamefont {T.}~\bibnamefont {Kippenberg}},\ }\bibfield
   {title} {\enquote {\bibinfo {title} {A maser based on dynamical backaction
  on microwave light},}\ }\href@noop {} {\bibfield  {journal} {\bibinfo
  {journal} {Physics Letters A}\ }\textbf {\bibinfo {volume} {382}},\ \bibinfo
  {pages} {2233--2237} (\bibinfo {year} {2018})}\BibitemShut {NoStop}%
\bibitem [{\citenamefont {Buczek}, \citenamefont {Freiberg},\ and\
  \citenamefont {Skolnick}(1973)}]{buczek1973laser}%
  \BibitemOpen
  \bibfield  {author} {\bibinfo {author} {\bibfnamefont {C.~J.}\ \bibnamefont
  {Buczek}}, \bibinfo {author} {\bibfnamefont {R.~J.}\ \bibnamefont
  {Freiberg}}, \ and\ \bibinfo {author} {\bibfnamefont {M.}~\bibnamefont
  {Skolnick}},\ }\bibfield  {title} {\enquote {\bibinfo {title} {Laser
  injection locking},}\ }\href@noop {} {\bibfield  {journal} {\bibinfo
  {journal} {Proceedings of the IEEE}\ }\textbf {\bibinfo {volume} {61}},\
  \bibinfo {pages} {1411--1431} (\bibinfo {year} {1973})}\BibitemShut {NoStop}%
\bibitem [{\citenamefont {Yamamoto}\ \emph {et~al.}(2008)\citenamefont
  {Yamamoto}, \citenamefont {Inomata}, \citenamefont {Watanabe}, \citenamefont
  {Matsuba}, \citenamefont {Miyazaki}, \citenamefont {Oliver}, \citenamefont
  {Nakamura},\ and\ \citenamefont {Tsai}}]{yamamoto2008flux}%
  \BibitemOpen
  \bibfield  {author} {\bibinfo {author} {\bibfnamefont {T.}~\bibnamefont
  {Yamamoto}}, \bibinfo {author} {\bibfnamefont {K.}~\bibnamefont {Inomata}},
  \bibinfo {author} {\bibfnamefont {M.}~\bibnamefont {Watanabe}}, \bibinfo
  {author} {\bibfnamefont {K.}~\bibnamefont {Matsuba}}, \bibinfo {author}
  {\bibfnamefont {T.}~\bibnamefont {Miyazaki}}, \bibinfo {author}
  {\bibfnamefont {W.~D.}\ \bibnamefont {Oliver}}, \bibinfo {author}
  {\bibfnamefont {Y.}~\bibnamefont {Nakamura}}, \ and\ \bibinfo {author}
  {\bibfnamefont {J.}~\bibnamefont {Tsai}},\ }\bibfield  {title} {\enquote
  {\bibinfo {title} {Flux-driven josephson parametric amplifier},}\ }\href@noop
  {} {\bibfield  {journal} {\bibinfo  {journal} {Applied Physics Letters}\
  }\textbf {\bibinfo {volume} {93}},\ \bibinfo {pages} {042510} (\bibinfo
  {year} {2008})}\BibitemShut {NoStop}%
\bibitem [{\citenamefont {Roy}\ and\ \citenamefont
  {Devoret}(2016)}]{roy2016introduction}%
  \BibitemOpen
  \bibfield  {author} {\bibinfo {author} {\bibfnamefont {A.}~\bibnamefont
  {Roy}}\ and\ \bibinfo {author} {\bibfnamefont {M.}~\bibnamefont {Devoret}},\
  }\bibfield  {title} {\enquote {\bibinfo {title} {Introduction to parametric
  amplification of quantum signals with josephson circuits},}\ }\href@noop {}
  {\bibfield  {journal} {\bibinfo  {journal} {Comptes Rendus Physique}\
  }\textbf {\bibinfo {volume} {17}},\ \bibinfo {pages} {740--755} (\bibinfo
  {year} {2016})}\BibitemShut {NoStop}%
\bibitem [{\citenamefont {Krantz}\ \emph {et~al.}(2013)\citenamefont {Krantz},
  \citenamefont {Reshitnyk}, \citenamefont {Wustmann}, \citenamefont
  {Bylander}, \citenamefont {Gustavsson}, \citenamefont {Oliver}, \citenamefont
  {Duty}, \citenamefont {Shumeiko},\ and\ \citenamefont
  {Delsing}}]{krantz2013investigation}%
  \BibitemOpen
  \bibfield  {author} {\bibinfo {author} {\bibfnamefont {P.}~\bibnamefont
  {Krantz}}, \bibinfo {author} {\bibfnamefont {Y.}~\bibnamefont {Reshitnyk}},
  \bibinfo {author} {\bibfnamefont {W.}~\bibnamefont {Wustmann}}, \bibinfo
  {author} {\bibfnamefont {J.}~\bibnamefont {Bylander}}, \bibinfo {author}
  {\bibfnamefont {S.}~\bibnamefont {Gustavsson}}, \bibinfo {author}
  {\bibfnamefont {W.~D.}\ \bibnamefont {Oliver}}, \bibinfo {author}
  {\bibfnamefont {T.}~\bibnamefont {Duty}}, \bibinfo {author} {\bibfnamefont
  {V.}~\bibnamefont {Shumeiko}}, \ and\ \bibinfo {author} {\bibfnamefont
  {P.}~\bibnamefont {Delsing}},\ }\bibfield  {title} {\enquote {\bibinfo
  {title} {Investigation of nonlinear effects in josephson parametric
  oscillators used in circuit quantum electrodynamics},}\ }\href@noop {}
  {\bibfield  {journal} {\bibinfo  {journal} {New Journal of Physics}\ }\textbf
  {\bibinfo {volume} {15}},\ \bibinfo {pages} {105002} (\bibinfo {year}
  {2013})}\BibitemShut {NoStop}%
\bibitem [{\citenamefont {Wustmann}\ and\ \citenamefont
  {Shumeiko}(2019)}]{wustmann2019parametric}%
  \BibitemOpen
  \bibfield  {author} {\bibinfo {author} {\bibfnamefont {W.}~\bibnamefont
  {Wustmann}}\ and\ \bibinfo {author} {\bibfnamefont {V.}~\bibnamefont
  {Shumeiko}},\ }\bibfield  {title} {\enquote {\bibinfo {title} {Parametric
  effects in circuit quantum electrodynamics},}\ }\href@noop {} {\bibfield
  {journal} {\bibinfo  {journal} {Low Temperature Physics}\ }\textbf {\bibinfo
  {volume} {45}},\ \bibinfo {pages} {848--869} (\bibinfo {year}
  {2019})}\BibitemShut {NoStop}%
\bibitem [{\citenamefont {Dykman}\ \emph {et~al.}(1998)\citenamefont {Dykman},
  \citenamefont {Maloney}, \citenamefont {Smelyanskiy},\ and\ \citenamefont
  {Silverstein}}]{dykman1998fluctuational_phaseflip}%
  \BibitemOpen
  \bibfield  {author} {\bibinfo {author} {\bibfnamefont {M.}~\bibnamefont
  {Dykman}}, \bibinfo {author} {\bibfnamefont {C.}~\bibnamefont {Maloney}},
  \bibinfo {author} {\bibfnamefont {V.}~\bibnamefont {Smelyanskiy}}, \ and\
  \bibinfo {author} {\bibfnamefont {M.}~\bibnamefont {Silverstein}},\
  }\bibfield  {title} {\enquote {\bibinfo {title} {Fluctuational phase-flip
  transitions in parametrically driven oscillators},}\ }\href@noop {}
  {\bibfield  {journal} {\bibinfo  {journal} {Physical Review E}\ }\textbf
  {\bibinfo {volume} {57}},\ \bibinfo {pages} {5202} (\bibinfo {year}
  {1998})}\BibitemShut {NoStop}%
\bibitem [{\citenamefont {Santoso}\ \emph {et~al.}(2012)\citenamefont
  {Santoso}, \citenamefont {McGranaghan}, \citenamefont {Dugan},\ and\
  \citenamefont {Beaty}}]{santoso2012electrical_spur}%
  \BibitemOpen
  \bibfield  {author} {\bibinfo {author} {\bibfnamefont {S.}~\bibnamefont
  {Santoso}}, \bibinfo {author} {\bibfnamefont {M.~F.}\ \bibnamefont
  {McGranaghan}}, \bibinfo {author} {\bibfnamefont {R.~C.}\ \bibnamefont
  {Dugan}}, \ and\ \bibinfo {author} {\bibfnamefont {H.~W.}\ \bibnamefont
  {Beaty}},\ }\href@noop {} {\emph {\bibinfo {title} {Electrical power systems
  quality}}}\ (\bibinfo  {publisher} {McGraw-Hill Education},\ \bibinfo {year}
  {2012})\BibitemShut {NoStop}%
\bibitem [{\citenamefont {Gao}\ \emph {et~al.}(2007)\citenamefont {Gao},
  \citenamefont {Zmuidzinas}, \citenamefont {Mazin}, \citenamefont {LeDuc},\
  and\ \citenamefont {Day}}]{gao2007noise}%
  \BibitemOpen
  \bibfield  {author} {\bibinfo {author} {\bibfnamefont {J.}~\bibnamefont
  {Gao}}, \bibinfo {author} {\bibfnamefont {J.}~\bibnamefont {Zmuidzinas}},
  \bibinfo {author} {\bibfnamefont {B.~A.}\ \bibnamefont {Mazin}}, \bibinfo
  {author} {\bibfnamefont {H.~G.}\ \bibnamefont {LeDuc}}, \ and\ \bibinfo
  {author} {\bibfnamefont {P.~K.}\ \bibnamefont {Day}},\ }\bibfield  {title}
  {\enquote {\bibinfo {title} {Noise properties of superconducting coplanar
  waveguide microwave resonators},}\ }\href@noop {} {\bibfield  {journal}
  {\bibinfo  {journal} {Applied Physics Letters}\ }\textbf {\bibinfo {volume}
  {90}},\ \bibinfo {pages} {102507} (\bibinfo {year} {2007})}\BibitemShut
  {NoStop}%
\bibitem [{\citenamefont {Machlup}(1954)}]{machlup1954noise_RTS}%
  \BibitemOpen
  \bibfield  {author} {\bibinfo {author} {\bibfnamefont {S.}~\bibnamefont
  {Machlup}},\ }\bibfield  {title} {\enquote {\bibinfo {title} {Noise in
  semiconductors: spectrum of a two-parameter random signal},}\ }\href@noop {}
  {\bibfield  {journal} {\bibinfo  {journal} {Journal of Applied Physics}\
  }\textbf {\bibinfo {volume} {25}},\ \bibinfo {pages} {341--343} (\bibinfo
  {year} {1954})}\BibitemShut {NoStop}%
\bibitem [{\citenamefont {Yuzhelevski}, \citenamefont {Yuzhelevski},\ and\
  \citenamefont {Jung}(2000)}]{yuzhelevski2000random_RTS}%
  \BibitemOpen
  \bibfield  {author} {\bibinfo {author} {\bibfnamefont {Y.}~\bibnamefont
  {Yuzhelevski}}, \bibinfo {author} {\bibfnamefont {M.}~\bibnamefont
  {Yuzhelevski}}, \ and\ \bibinfo {author} {\bibfnamefont {G.}~\bibnamefont
  {Jung}},\ }\bibfield  {title} {\enquote {\bibinfo {title} {Random telegraph
  noise analysis in time domain},}\ }\href@noop {} {\bibfield  {journal}
  {\bibinfo  {journal} {Review of Scientific Instruments}\ }\textbf {\bibinfo
  {volume} {71}},\ \bibinfo {pages} {1681--1688} (\bibinfo {year}
  {2000})}\BibitemShut {NoStop}%
\end{thebibliography}%


\providecommand{\noopsort}[1]{}\providecommand{\singleletter}[1]{#1}%
%


\begin{thebibliography}{5}%
\makeatletter
\providecommand \@ifxundefined [1]{%
 \@ifx{#1\undefined}
}%
\providecommand \@ifnum [1]{%
 \ifnum #1\expandafter \@firstoftwo
 \else \expandafter \@secondoftwo
 \fi
}%
\providecommand \@ifx [1]{%
 \ifx #1\expandafter \@firstoftwo
 \else \expandafter \@secondoftwo
 \fi
}%
\providecommand \natexlab [1]{#1}%
\providecommand \enquote  [1]{``#1''}%
\providecommand \bibnamefont  [1]{#1}%
\providecommand \bibfnamefont [1]{#1}%
\providecommand \citenamefont [1]{#1}%
\providecommand \href@noop [0]{\@secondoftwo}%
\providecommand \href [0]{\begingroup \@sanitize@url \@href}%
\providecommand \@href[1]{\@@startlink{#1}\@@href}%
\providecommand \@@href[1]{\endgroup#1\@@endlink}%
\providecommand \@sanitize@url [0]{\catcode `\\12\catcode `\$12\catcode
  `\&12\catcode `\#12\catcode `\^12\catcode `\_12\catcode `\%12\relax}%
\providecommand \@@startlink[1]{}%
\providecommand \@@endlink[0]{}%
\providecommand \url  [0]{\begingroup\@sanitize@url \@url }%
\providecommand \@url [1]{\endgroup\@href {#1}{\urlprefix }}%
\providecommand \urlprefix  [0]{URL }%
\providecommand \Eprint [0]{\href }%
\providecommand \doibase [0]{https://doi.org/}%
\providecommand \selectlanguage [0]{\@gobble}%
\providecommand \bibinfo  [0]{\@secondoftwo}%
\providecommand \bibfield  [0]{\@secondoftwo}%
\providecommand \translation [1]{[#1]}%
\providecommand \BibitemOpen [0]{}%
\providecommand \bibitemStop [0]{}%
\providecommand \bibitemNoStop [0]{.\EOS\space}%
\providecommand \EOS [0]{\spacefactor3000\relax}%
\providecommand \BibitemShut  [1]{\csname bibitem#1\endcsname}%
\let\auto@bib@innerbib\@empty
\bibitem [{\citenamefont {Clerk}\ \emph {et~al.}(2010)\citenamefont {Clerk},
  \citenamefont {Devoret}, \citenamefont {Girvin}, \citenamefont {Marquardt},\
  and\ \citenamefont {Schoelkopf}}]{clerk2010introduction}%
  \BibitemOpen
  \bibfield  {author} {\bibinfo {author} {\bibfnamefont {A.~A.}\ \bibnamefont
  {Clerk}}, \bibinfo {author} {\bibfnamefont {M.~H.}\ \bibnamefont {Devoret}},
  \bibinfo {author} {\bibfnamefont {S.~M.}\ \bibnamefont {Girvin}}, \bibinfo
  {author} {\bibfnamefont {F.}~\bibnamefont {Marquardt}},\ and\ \bibinfo
  {author} {\bibfnamefont {R.~J.}\ \bibnamefont {Schoelkopf}},\ }\href@noop {}
  {\bibfield  {journal} {\bibinfo  {journal} {Reviews of Modern Physics}\
  }\textbf {\bibinfo {volume} {82}},\ \bibinfo {pages} {1155} (\bibinfo {year}
  {2010})}\BibitemShut {NoStop}%
\bibitem [{\citenamefont {Lin}\ \emph {et~al.}(2014)\citenamefont {Lin},
  \citenamefont {Inomata}, \citenamefont {Koshino}, \citenamefont {Oliver},
  \citenamefont {Nakamura}, \citenamefont {Tsai},\ and\ \citenamefont
  {Yamamoto}}]{lin2014josephsonS}%
  \BibitemOpen
  \bibfield  {author} {\bibinfo {author} {\bibfnamefont {Z.}~\bibnamefont
  {Lin}}, \bibinfo {author} {\bibfnamefont {K.}~\bibnamefont {Inomata}},
  \bibinfo {author} {\bibfnamefont {K.}~\bibnamefont {Koshino}}, \bibinfo
  {author} {\bibfnamefont {W.}~\bibnamefont {Oliver}}, \bibinfo {author}
  {\bibfnamefont {Y.}~\bibnamefont {Nakamura}}, \bibinfo {author}
  {\bibfnamefont {J.-S.}\ \bibnamefont {Tsai}},\ and\ \bibinfo {author}
  {\bibfnamefont {T.}~\bibnamefont {Yamamoto}},\ }\href@noop {} {\bibfield
  {journal} {\bibinfo  {journal} {Nature communications}\ }\textbf {\bibinfo
  {volume} {5}},\ \bibinfo {pages} {1} (\bibinfo {year} {2014})}\BibitemShut
  {NoStop}%
\bibitem [{\citenamefont {Gao}\ \emph {et~al.}(2007)\citenamefont {Gao},
  \citenamefont {Zmuidzinas}, \citenamefont {Mazin}, \citenamefont {LeDuc},\
  and\ \citenamefont {Day}}]{gao2007noiseS}%
  \BibitemOpen
  \bibfield  {author} {\bibinfo {author} {\bibfnamefont {J.}~\bibnamefont
  {Gao}}, \bibinfo {author} {\bibfnamefont {J.}~\bibnamefont {Zmuidzinas}},
  \bibinfo {author} {\bibfnamefont {B.~A.}\ \bibnamefont {Mazin}}, \bibinfo
  {author} {\bibfnamefont {H.~G.}\ \bibnamefont {LeDuc}},\ and\ \bibinfo
  {author} {\bibfnamefont {P.~K.}\ \bibnamefont {Day}},\ }\href@noop {}
  {\bibfield  {journal} {\bibinfo  {journal} {Applied Physics Letters}\
  }\textbf {\bibinfo {volume} {90}},\ \bibinfo {pages} {102507} (\bibinfo
  {year} {2007})}\BibitemShut {NoStop}%
\bibitem [{\citenamefont {Bhai}\ \emph {et~al.}(2022)\citenamefont {Bhai},
  \citenamefont {Mukai}, \citenamefont {Yamamoto},\ and\ \citenamefont
  {Tsai}}]{bhai2022noiseS}%
  \BibitemOpen
  \bibfield  {author} {\bibinfo {author} {\bibfnamefont {G.~L.}\ \bibnamefont
  {Bhai}}, \bibinfo {author} {\bibfnamefont {H.}~\bibnamefont {Mukai}},
  \bibinfo {author} {\bibfnamefont {T.}~\bibnamefont {Yamamoto}},\ and\
  \bibinfo {author} {\bibfnamefont {J.-S.}\ \bibnamefont {Tsai}},\ }\href@noop
  {} {\bibfield  {journal} {\bibinfo  {journal} {arXiv preprint
  arXiv:2210.15116}\ } (\bibinfo {year} {2022})}\BibitemShut {NoStop}%
\bibitem [{\citenamefont {Welch}(1967)}]{welch1967use}%
  \BibitemOpen
  \bibfield  {author} {\bibinfo {author} {\bibfnamefont {P.}~\bibnamefont
  {Welch}},\ }\href@noop {} {\bibfield  {journal} {\bibinfo  {journal} {IEEE
  Transactions on audio and electroacoustics}\ }\textbf {\bibinfo {volume}
  {15}},\ \bibinfo {pages} {70} (\bibinfo {year} {1967})}\BibitemShut {NoStop}%
\end{thebibliography}%

\end{document}



\title{Supplemental Material: \\
Mitigation of noise in Josephson parametric oscillator by injection locking}



\author{Gopika Lakshmi Bhai}\email{gopika.lakshmibhai@gmail.com}
\affiliation{%
Graduate School of Science, Tokyo University of Science,1–3 Kagurazaka, Shinjuku, Tokyo 162–0825, Japan
}%

\affiliation{%
Research Institute for Science and Technology, Tokyo University of Science, 1-3 Kagurazaka, Shinjuku-ku, Tokyo 162-8601, Japan
}%

\affiliation{%
RIKEN Center for Quantum Computing (RQC), 2–1 Hirosawa, Wako, Saitama 351–0198, Japan
}%

\author{Hiroto Mukai}%
\affiliation{%
RIKEN Center for Quantum Computing (RQC), 2–1 Hirosawa, Wako, Saitama 351–0198, Japan
}%

\author{ Jaw-Shen Tsai}\email{tsai@riken.jp}
\affiliation{%
Graduate School of Science, Tokyo University of Science,1–3 Kagurazaka, Shinjuku, Tokyo 162–0825, Japan
}%

\affiliation{%
Research Institute for Science and Technology, Tokyo University of Science, 1-3 Kagurazaka, Shinjuku-ku, Tokyo 162-8601, Japan
}%

\affiliation{%
RIKEN Center for Quantum Computing (RQC), 2–1 Hirosawa, Wako, Saitama 351–0198, Japan
}%

\date{\today}


\maketitle

\renewcommand{\thesection}{S1}
\section{Device and Measurement Setup}
\label{measurement_setup}

\renewcommand{\thefigure}{S1}
\begin{figure*}[ht]
\begin{center}
\includegraphics[keepaspectratio]{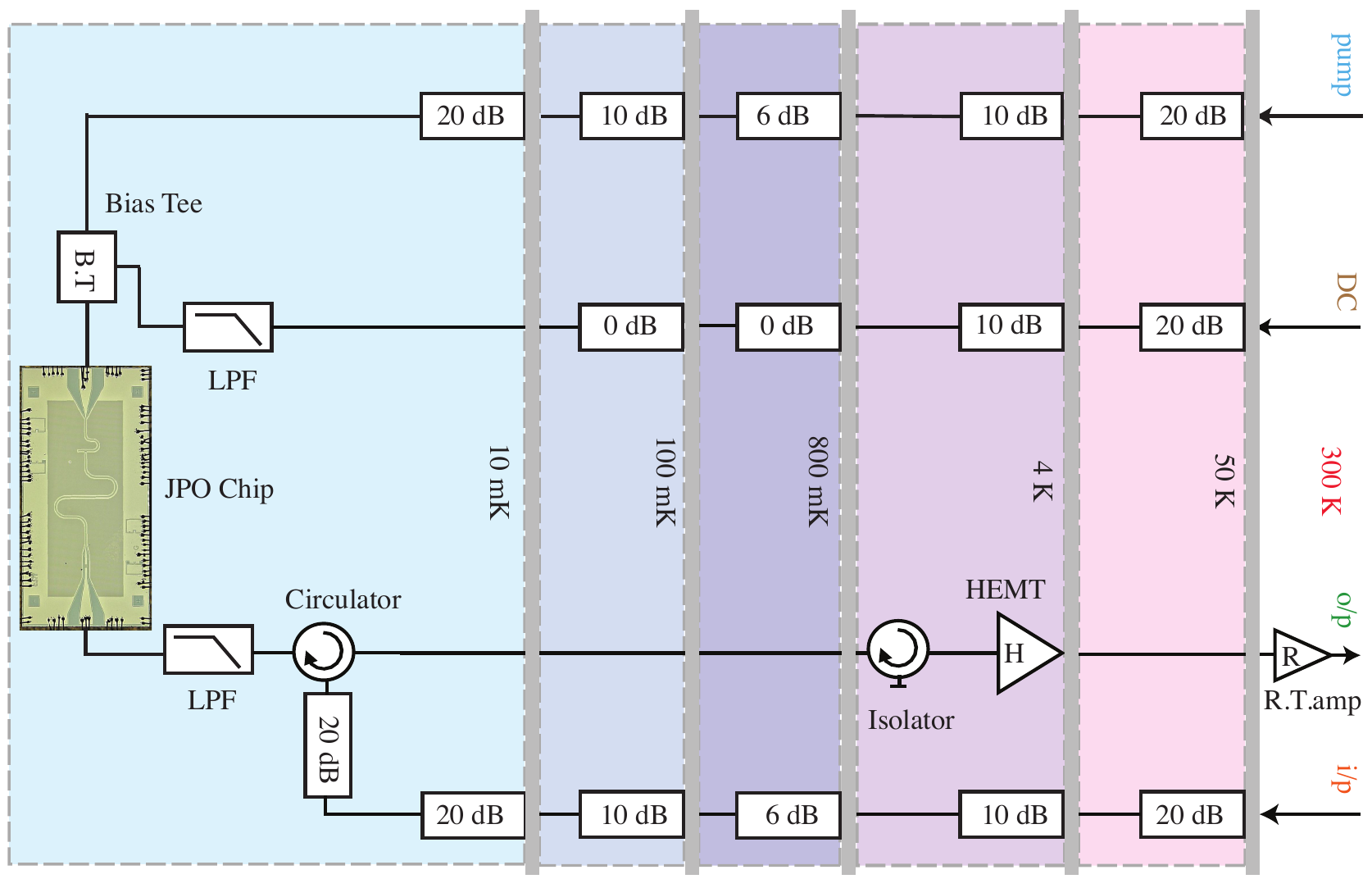} 
\end{center}
\caption{ 
Schematic of the cryogenic circuit. JPO chip is mounted at the base temperature of the dilution refrigerator and cooled down to 10 mK. DC flux bias is applied through the pump port using a bias tee. Input and output signals are directed using a circulator. The output signal is amplified with HEMT at 4K and further amplified using a room-temperature amplifier.}
\end{figure*}

The device consisting of a quarter wavelength resonator terminated with the dc-SQUID is cooled down to 10 mK using a dilution refrigerator. The cryogenic circuitry is illustrated in Fig.~S1. In order to characterize the device, we measure the reflection coefficient of the resonator cavity $S_{11}$ as a function of frequency. In the absence of flux bias, the bare resonator frequency is measured to be 6.2 GHz. The dc-SQUID provides flux tunable nonlinear inductance, and by varying the dc flux applied through an on-chip flux bias line, we examine the flux frequency curve. Our phase noise measurements are carried out at a fixed bias $\Phi_{\rm{dc}}/\Phi_{0} = 0.3$, where the resonator frequency is measured to be $\omega_{s}/2\pi = 5.95$ GHz. By fitting the complex reflection coefficient $S_{11}$ measured at this fixed bias, we estimate the internal and external cavity loss rates of the resonator, which are found to be $\kappa_{\rm{ext}}/2\pi = 17$ MHz and $\kappa_{\rm{int}}/2\pi = 0.3$ MHz. Furthermore, in order to parametrically drive the resonator, we apply a pump tone at $2\omega_{\rm{s}}$ 
 to periodically modulate the dc-SQUID inductance. Thus, by the periodic variation of the resonator frequency around  $2\omega_{\rm{s}}$, self-sustained oscillation builds up in the resonator when the drive strength $P_{\rm{p}}$ crosses the parametric threshold. The phase noise measurements of the JPO operating above the threshold are carried out at a fixed pump power $P_{\rm{p}} = -56$~dBm and with a varying input injection locking signal (ILS) power. The number of photons inside the cavity is calculated using the relation $N_{\rm{p}} = 4 P_{\rm{s}} \kappa_{\rm{ext}}/ \hbar \omega_{\rm{s}}\kappa_{\rm{tot}}^2$, where $\kappa_{\rm{tot}} = \kappa_{\rm{int}} + \kappa_{\rm{ext}} $, $P_{\rm{s}}$ and  $\omega_{\rm{s}}$ are the power and frequency of the ILS~\cite{clerk2010introduction,lin2014josephsonS}. 

\renewcommand{\thesection}{S2}
\section{Noise analysis method}
\label{noise analysis}

For the characterization of noise, we use the microwave homodyne interferometric measurement setup, as shown in Fig.~2 in the main text. The phase-coherent microwave synthesizers generate the frequency at $\omega_{\rm{s}}$, which is used to feed as the local oscillator at the same frequency $\omega_{\rm{s}}$ as the ILS, and the pump tone by doubling the frequency $\omega_{\rm{p}} = 2 \omega_{\rm{s}}$. The output from the JPO is compared with the local oscillator at the same frequency using an IQ mixer. The output I and Q voltages which are proportional to the in-phase and quadrature amplitude, are digitized with a 14-bit Keysight Digitizer with an onboard field programmable gate array (FPGA). We record the unbreakable real-time data for 10 seconds with a sampling rate of 1MSa/s and analyze the time domain data recorded at a single shot to extract the noise in phase and amplitude quadratures. 
\\

We first calculate the fluctuation in I and Q about its mean, $\delta \eta(t) = [\delta I(t), \delta Q(t)]^{T}$. In order to quantify the noise data $\delta \eta(t)$, we construct the spectral domain noise covariance matrix $S_{\nu}$ defined by~\cite{gao2007noiseS,bhai2022noiseS},

\renewcommand{\theequation}{S\arabic{equation}}
\begin{align} 
\left\langle\delta \eta(\nu) \delta \eta^{\dagger}\left(\nu^{\prime}\right)\right\rangle &= S(\nu) \delta\left(\nu-\nu^{\prime}\right), 
S(\nu) =\left(\begin{array}{cc}S_{I I}(\nu) & S_{I Q}(\nu) \\ S_{I Q}^{*}(\nu) & S_{Q Q}(\nu)\end{array}\right).
\end{align}

$\delta \eta(\nu)$ and $\delta \eta(\nu)^\dagger$ are the Fourier transform of the time domain data and its Hermitian conjugate. The diagonal elements of the matrix $S_{\nu}$ are the autocorrelation power spectra, and the off-diagonal elements  are the cross-correlation power spectra calculated using Welche's method~\cite{welch1967use}. We note that the star on the off-diagonal element represents the complex conjugate. To extract the phase and amplitude noise, we diagonalize the matrix using a unitary transformation with an ordinary rotation applied to the real part of $S_{\nu}$, 

\begin{equation}
O^{T}(\nu) \operatorname{Re} S(\nu) O(\nu)=\left(\begin{array}{cc}S_{a a}(\nu) & 0 \\ 0 & S_{b b}(\nu)\end{array}\right).
\end{equation}

Here $S_{a a}$ and $S_{b b}$ quantify the noise spectral density in the phase and amplitude quadrature. In our study, we investigate the effect of the ILS on the phase noise power spectral density $S_{a a}$ characterized for the JPO operating above the threshold by applying a weak ILS and by varying the power of ILS.


\bibliography{ref_suppl}